\documentclass[authoryear,letterpaper,preprint]{elsarticle}

\usepackage{subfigure}
\usepackage{amsmath}
\usepackage{amssymb}
\usepackage{graphicx}

\journal{Mechanics of Materials}

\begin{document}

\begin{frontmatter}
\title{{Micro-mechanics of Fabric and Failure\\ in Granular Materials}}
\author[mrk]{Matthew R. Kuhn}
\ead{kuhn@up.edu}
\address[mrk]{University of Portland, 5000 N. Willamette Blvd., 
              Portland, OR, 97203~USA, 
              Tel.~(1)-503-943-7361,~Fax~(1)-503-943-7316.}
\begin{abstract}
The Paper addresses the underlying source of
two forms of induced anisotropy in granular materials: contact orientation
anisotropy and contact force an\-iso\-tropy. A rational, mathematical
structure is reviewed for the manner in which fabric anisotropy emerges
and evolves during loading. Fabric is expressed as an orientation
density, and transport phenomena such as convection, contact generation,
and diffusion control the rate of fabric evolution during loading.
The Paper proposes specific measurable forms for all terms, based
upon the micro-mechanics of particle interactions. Discrete element
(DEM) simulations are used to verify and quantify these terms, so
that the theory can be applied to general loading conditions. The
DEM simulations are of densely packed durable spheres, and the emphasis
is on soil behavior at large strains, specifically on fabric and
strength at the critical state. Once the theory has been developed
and quantified, it is applied to predict the effect of the intermediate
principal stress on strength.
\end{abstract}
\begin{keyword}
Fabric/structure of soils; constitutive
relations; failure; numerical modelling and analysis; shear strength.
\end{keyword}
\end{frontmatter}
\section{Introduction}
A granular material, such as a soil, possesses an inherent
fabric that is imprinted
during its formation. This initial fabric usually imparts a directional,
an\-iso\-tropic character, so that strength and stiffness depend upon
the direction of loading relative to the original deposition 
\citep{Arthur:1972a}.
\citet{Oda:1972a} defined fabric as the spatial arrangement of particles
and voids, and he showed that the microscopic fabric
can exhibit two forms of anisotropy: a preferred orientation of elongated
or flat particles, and a prevalence of inter-particle contacts in
preferred directions. In pioneering experiments, he also found
that the subsequent loading and deformation
of a soil alters the arrangement of its
particles and causes contacts to become increasingly aligned in the
direction of the major principal stress~---~a phenomenon now termed
\emph{stress-induced fabric anisotropy }\citep{Oda:1972b,Oda:1972c}\emph{.}
These early studies, which derived from a geometric view of fabric,
were later augmented to include a kinetic or statical aspect of anisotropy.
Photo-elastic experiments have shown that forces between particles
become largest at contacts that are aligned with the loading direction;
moreover, Cundall and others have shown that deviatoric stress is
largely an expression of such \emph{stress-induced force anisotropy}
\citep{Cundall:1983a,Thornton:1986a}.
\par
The current study addresses the induced anisotropies of contact orientation
and contact force. The Paper develops a systematic means of tracking
and analyzing the evolution of anisotropy. That is, the focus is on
the underlying source and rate of induced anisotropy~--- on the rate of its
evolution rather than its state at a particular instant. 
Section~2 begins with a brief presentation 
conventional measures of fabric and stress and
then presents a rational, mathematical approach to the
manner in which fabric anisotropy emerges and evolves during loading.
The basis is an idea recently proposed by \citet{Ma:2006a}, and the
Paper goes beyond this original idea by supplying additional terms
and developing expressions for all terms. Although the first part
of the paper is primarily analytical, in Section~\ref{sec:Quantifying}
the Paper proposes specific forms for the various mathematical terms
and uses discrete element (DEM) simulations to verify and quantify
these terms. 
The DEM simulations are of densely packed durable spheres,
and the data analysis concentrates on soil behavior at large, failure
strains, specifically on fabric and strength at the critical state.
Once the theory has been developed and quantified, 
Section~\ref{sec:pi_plane}
applies the theory to predict the effect of the intermediate principal
stress on soil strength at the critical state. 
The theory provides
an explanation for the shape of the critical state yield surface,
and the predictions are compared with published results and with DEM
simulations.
\section{\label{sec:Theory}Fabric rate equations}
\begin{figure}
\centering
\includegraphics{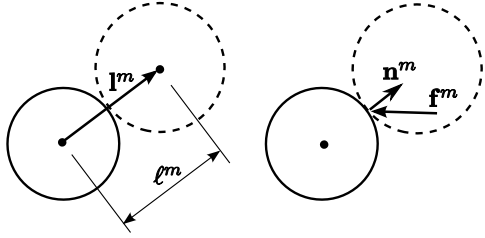}
\caption{Two particles at a contact $m$.}
\label{fig:particles}
\end{figure}

The Paper pursues two anisotropies, of contact orientation and of
contact force, and models their evolution during bulk loading. Fabric
anisotropy will be expressed in terms of the \emph{contact density}
$\widehat{g}(\mathbf{n})$, a function that describes contact orientation
within an assembly of particles. The individual orientation $\mathbf{n}^{m}$
of a single $m^{\mathrm{th}}$ contact is the unit vector normal
to the surfaces of two particles at their contact point (Fig.~\ref{fig:particles}).%
The density $\widehat{g}(\mathbf{n})$ within an entire assembly is
the average number of contacts having a common orientation $\mathbf{n}$,
as expressed per particle and per unit of area on the unit sphere
(i.e., per steradian of solid angle). If a granular assembly contains
$N$ particles having $M$ contacts, then the bulk average of contacts
per particle, $M/N$, is the integral of density $\widehat{g}(\mathbf{n})$
across the unit sphere,\[
M/N=\int_{\Omega}\widehat{g}(\mathbf{n})\, d\Omega=4\pi\widehat{g_{\text{avg}}}\]
where $\Omega$ is the unit sphere surface and $\widehat{g_{\text{avg}}}$
is the average density. Note that the more commonly used \emph{average
coordination number} is simply twice $M/N$. An initially isotropic
particle arrangement, with no preferred direction of contact orientation,
has a uniform density $\widehat{g}(\mathbf{n})=M/N/(4\pi)$ across
all of $\Omega$, measured in contacts per particle per steradian
of solid angle. Upon loading, the density becomes anisotropic, with
larger values in the direction of compressive loading.
\par
In a similar manner, we can consider a {}``$\:\widehat{\:}\:$''
\emph{density of contact force}, $\widehat{\mathbf{f}}(\mathbf{n})$, also
a function of contact orientation $\mathbf{n}$. This vector density
has units of force per particle per steradian on the unit sphere.
The \emph{average force} among those contacts having a particular
orientation $\mathbf{n}$ will be written as $\overline{\mathbf{f}}(\mathbf{n})$
and is simply the force density divided by the contact density:
\begin{equation}
\overline{\mathbf{f}}(\mathbf{n})=\widehat{\mathbf{f}}(\mathbf{n})\,/\,\widehat{g}(\mathbf{n})
\label{eq:avg_f}
\end{equation}
\par
To relate these densities to stress, we begin with the Cauchy formula
for the average stress $\boldsymbol{\sigma}$ within a granular region\[
\boldsymbol{\sigma}=\frac{1}{V}\sum_{m=1}^{M}\mathbf{l}^{m}\otimes\mathbf{f}^{m}\]
a discrete sum of dyadic products $\mathbf{l}^{m}\otimes\mathbf{f}^{m}$
(= $l_{i}^{m}f_{j}^{m}$) for the $M$ contacts within the region's
volume $V$. Vector $\mathbf{f}^{m}$ is the contact force, and $\mathbf{l}^{m}$
is the branch vector between the centers of two contacting particles
(Fig.~\ref{fig:particles}). For spherical particles, vector $\mathbf{l}^{m}$
is the product $\ell^{m}\mathbf{n}^{m}$ of the branch length $\ell^{m}$
and the contact's unit normal vector $\mathbf{n}^{m}$, so that
\begin{equation}
\boldsymbol{\sigma}=\frac{1}{V}\sum_{m=1}^{M}\ell^{m}\mathbf{n}^{m}\otimes\mathbf{f}^{m}\approx\frac{\overline{\ell}}{V}\sum_{m=1}^{M}\mathbf{n}^{m}\otimes\mathbf{f}^{m}
\label{eq:stress2}
\end{equation}
where the approximation includes the average branch length $\overline{\ell}$
among all contacts. In making this approximation, we ignore the small
correlation between individual lengths $\ell^{m}$ and the corresponding
contact directions and forces. Noting that $\widehat{\mathbf{f}}(\mathbf{n})$
is the density of contact force per particle, the average stress within
a sphere assembly can be expressed as
\begin{equation}
\boldsymbol{\sigma}\approx\frac{\overline{\ell}N}{V}\int_{\Omega}\mathbf{n}\otimes\widehat{\mathbf{f}}(\mathbf{n})\, d\Omega=\frac{\overline{\ell}}{\overline{v}}\int_{\Omega}\mathbf{n}\otimes\widehat{\mathbf{f}}(\mathbf{n})\, d\Omega
\label{eq:Stress3}
\end{equation}
with $\mathbf{n}\otimes\widehat{\mathbf{f}}(\mathbf{n})=n_{i}\widehat{f}_{j}(\mathbf{n})$.
In this form, we replace the discrete sum in Eq.~(\ref{eq:stress2})
with an integral of density on the unit sphere $\Omega$, and we introduce
$\overline{v}$, the average volume of a particle and its associated
void space ($\overline{v}=V/N$). Because the force density $\widehat{\mathbf{f}}(\mathbf{n})$
is the product $\widehat{g}(\mathbf{n})\overline{\mathbf{f}}(\mathbf{n})$,
deviatoric stress is the result of anisotropies in both contact orientation
$\widehat{g}(\mathbf{n})$ and average contact force $\overline{\mathbf{f}}(\mathbf{n})$
(see \citealp{Rothenburg:1989a}).
\par
The above principles apply to the \emph{state} of anisotropy at any instant
and are well established in the literature 
(for reviews, see \citealt{Oda:1999b,Nemat-Nasser:2001b}).
Equation~(\ref{eq:Stress3}) will henceforth be used to investigate 
the rates of fabric and stress evolution
during loading, the primary intent of the Paper. The stress
rate will depend, in part, upon the rate at which the contact force
density $\widehat{\mathbf{f}}(\mathbf{n})$ evolves, which will be
written as $\left.\partial\widehat{\mathbf{f}}(\mathbf{n},t)/\partial t\right|_{\mathbf{n}}$,
or simply $\partial\widehat{\mathbf{f}}(\mathbf{n})/\partial t$.
The partial derivatives emphasize that the rate of density $\widehat{\mathbf{f}}(\mathbf{n})$
is measured at fixed orientations $\mathbf{n}$, even though the motions
of individual particles will cause contacts to pass through any given
orientation. Focusing attention on small fixed portions $d\Omega$
of the unit sphere, the stress rate $\dot{\boldsymbol{\sigma}}$ is
computed from the rate at which force density $\widehat{\mathbf{f}}(\mathbf{n})$
changes within these regions:
\begin{equation}
\dot{\boldsymbol{\sigma}}=-\frac{\dot{v}}{\overline{v}}\boldsymbol{\sigma}+\frac{\overline{\ell}}{\overline{v}}\int_{\Omega}\mathbf{n}\otimes\left.\frac{\partial}{\partial t}\widehat{\mathbf{f}}(\mathbf{n})\right|_{\mathbf{n}}\, d\Omega
\label{eq:stress_rate1}
\end{equation}
again assuming spherical particles. This rate expression includes
a possible volume change during the loading process ($\dot{v}=d(V/N)/dt=d\overline{v}/dt$)
but assumes that the average branch length $\overline{\ell}$ remains
constant,
an assumption that is appropriate for hard particles having a ratio
of stress and shear modulus $\sigma / G$ that is small.
\par
\begin{figure}
\centering
\includegraphics{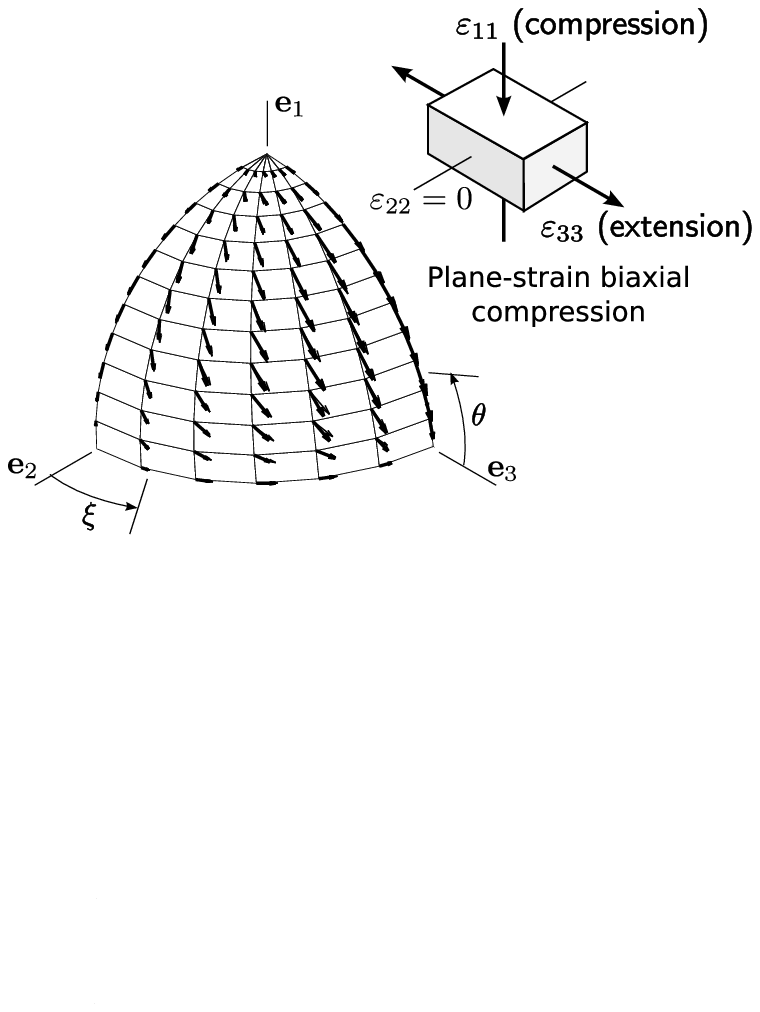}
\caption{Migration vectors $\dot{\mathbf{n}}$ and
$\dot{\mathbf{n}}^{\mathrm{proj}}$
on the unit sphere during plane-strain biaxial compression at the
critical state, from DEM simulations 
(Sections~\ref{section:DEM} and~\ref{section:Migration_rate}).}
\label{fig:migration}
\end{figure}

Particles will roll and slide across each other during loading, and
the orientation of any particular contact, say $\mathbf{n}^{m}$,
can shift to a new orientation $\mathbf{n}^{m}+\dot{\mathbf{n}}^{m}dt$.
These contact movements can be highly erratic, but when tracked by
the thousands, contacts are observed to migrate from directions of
bulk compression toward directions of extension. This average, prevailing
migration rate, denoted as $\dot{\mathbf{n}}(\mathbf{n})$, is a vector
field tangent to (and on) the unit sphere and has units of radians
per time. The migration $\dot{\mathbf{n}}(\mathbf{n})$ is a function
of orientation $\mathbf{n}$ and will depend upon the loading
process. As an example, Fig.~\ref{fig:migration} illustrates the
average migration vectors among over 800,000 contacts, as measured
in DEM simulations of sustained flow during biaxial plane-strain compression
at the critical (steady) state (see Sections~\ref{section:DEM}--\ref{section:Migration_rate}). %
As permitted by the symmetry of these loading conditions, the results
have been folded into a single octant of the unit sphere. Contacts
are seen to migrate (flow) from the compression direction, $\mathbf{n=\mathbf{e}}_{1}$,
toward the zero-strain direction $\mathbf{\mathbf{e}}_{2}$ and
toward the extension direction $\mathbf{\mathbf{e}}_{3}$. When
considered alone, this migration will be seen to have a softening
effect: by transporting contacts and their forces from the direction
of compression toward the direction of extension, migration usually diminishes
both fabric anisotropy and deviatoric stress. A functional form of
the migration $\dot{\mathbf{n}}(\mathbf{n})$ is proposed in Section~\ref{section:Migration_rate}.
\par
The stress evolution in Eq.~(\ref{eq:stress_rate1}) depends on the
rate of force density $\partial\widehat{\mathbf{f}}(\mathbf{n})/\partial t$,
a rate that will depend upon the interactions of the particles and
also upon the prevailing contact migration $\dot{\mathbf{n}}(\mathbf{n})$.
The density rate can be viewed as a transport problem on the surface
of the unit sphere. In this sense, the rate $\partial\widehat{\mathbf{f}}(\mathbf{n},t)/\partial t$
at a given, fixed orientation $\mathbf{n}$ is 
described by the Fokker-Planck equation with an additional source density:
\begin{equation}
\left.\frac{\partial\widehat{\mathbf{f}}(\mathbf{n})}{\partial t}\right|_{\mathbf{n}}=\frac{d\widehat{\mathbf{f}}(\mathbf{n})}{dt}\,-\boldsymbol{\nabla}_{\!_{\Omega}}\cdot\left(\dot{\mathbf{n}}\otimes\widehat{\mathbf{f}}(\mathbf{n})\right)+\left(\frac{\partial\widehat{\mathbf{f}}(\mathbf{n})}{\partial t}\right)_{\mathrm{diff}}
\label{eq:f_rate}
\end{equation}
where $\boldsymbol{\nabla}_{\!_{\Omega}}$ is the gradient $\partial(\bullet)/\partial\mathbf{n}$
on the unit sphere, and $\boldsymbol{\nabla}_{\!_{\Omega}}\cdot(\bullet)$
is the corresponding divergence, $\boldsymbol{\nabla}_{\!_{\Omega}}\cdot\mathbf{a}=a_{k,k}$.
This general form, except for the final (diffusion) term
and certain details among the other terms, was recently
proposed by \citet{Ma:2006a}. We describe the nature of each term
in the remainder of this section, noting that all are amenable to
rational analysis and to experimental measurement. In Section~\ref{sec:Quantifying},
we propose more specific forms for each term, verify these forms with
DEM data, and quantify the relevant material parameters that appear
within the specific forms. As a complement to the force rate in Eq.~(\ref{eq:f_rate}),
we must also consider the rate of contact density, which is also the
result of three terms:%
\begin{equation}
\left.\frac{\partial\widehat{g}(\mathbf{n})}{\partial t}\right|_{\mathbf{n}}=\frac{d\widehat{g}(\mathbf{n})}{dt}\,-\boldsymbol{\nabla}_{\!_{\Omega}}\cdot\left(\dot{\mathbf{n}}\,\widehat{g}(\mathbf{n})\right)+\left(\frac{\partial\widehat{g}(\mathbf{n})}{\partial t}\right)_{\mathrm{diff}}
\label{eq:g_rate}
\end{equation}
where the gradient $\boldsymbol{\nabla}_{\!_{\Omega}}\cdot(\dot{\mathbf{n}}\,\widehat{g})=(\dot{n}_{i}\widehat{g})_{,i}$.
\citet{Didwania:2001a} proposed an expression 
similar to Eq.~(\ref{eq:g_rate}) for the evolution
contact density, although the mean-field rotation of particle pairs was used
in place of the more general migration $\dot{\mathbf{n}}$.
\par
The first terms on the right of Eqs.~(\ref{eq:f_rate}) and~(\ref{eq:g_rate}), 
both $d\widehat{\mathbf{f}}/dt$ and
$d\widehat{g}/dt$, are \emph{source densities}, representing the
rates of force and contact density that are generated by the particle
interactions. The other terms account for convection and diffusion
processes which are associated with contact migration, as will be
discussed later. 
\par
Because contact density $\widehat{g}(\mathbf{n})$ is a scalar field,
its {}``$d\,$'' source density $d\widehat{g}(\mathbf{n})/dt$ is
somewhat easier to describe and measure than is the vector force rate
$d\widehat{\mathbf{f}}/dt$. Contacts are continually created and
broken during deviatoric loading, 
producing the net rate $d\widehat{g}(\mathbf{n})/dt$
at any given orientation $\mathbf{n}$. Unlike the force rate, the
contact rate is entirely a material rate, 
$d\widehat{g}(\mathbf{n})/dt=
(\partial\widehat{g}(\mathbf{n})/\partial t)_{\text{matl}}$,
and will be expressed as such in the remainder of the paper. As an
example, in our DEM simulations of dense particle packings, 4096 spherical
particles would typically touch at about 8190 contacts while the assembly
was flowing at the critical (steady) state in sustained biaxial plane-strain
compression (Section~\ref{section:DEM}). During such flow, over
6000 contacts were created and about the same number were broken with
each 1\% of continued deviatoric strain, yet maintaining the nearly
constant 8190 contacts at any instant. Contacts were predominantly
created in the compressive strain direction~--- at orientation $\mathbf{n=\mathbf{e}}_{1}$
in the simulations (Fig.~\ref{fig:migration})~--- and were predominantly
extinguished in the direction of extension, $\mathbf{n=\mathbf{e}}_{3}$.
This pattern of contact activity produces the net material rate and
is responsible for induced fabric anisotropy. In Section~\ref{section:contact_matl_rate},
we quantify the material rate, which is shown to be a function of
orientation $\mathbf{n}$, the loading conditions, and the material
characteristics of the particles.
\par
The source density of force, the $d\widehat{\mathbf{f}}/dt$ rate
in Eq.~(\ref{eq:f_rate}) is similar to the contact rate $d\widehat{g}/dt$
but also involves rotation of a vector field~--- a rotation that will
always accompany contact migration. For example, a single contact
force $\mathbf{f}^{m}$ between two particles will rotate as the
particles roll across each other, inducing a certain force increment
during the contact rotation $d\mathbf{n}^{m}$ (see \citealp{Kuhn:2005b}).
This \emph{induced rotational increment} must be added to any \emph{material
increment} in the force that might result from a changing indentation
at the contact:%
\begin{align}
\begin{split}d\mathbf{f}^{m}= & \left(d\mathbf{f}^{m}\right)_{\text{matl}}+\mathbf{f}^{m}\times(d\mathbf{n}^{m}\times\mathbf{n}^{m})
\\
 & +\frac{1}{2}\left[\left(d\boldsymbol{\theta}^{i}+d\boldsymbol{\theta}^{j}\right)\cdot\mathbf{n}^{m}\right]\left(\mathbf{n}^{m}\times\mathbf{f}^{m}\right)
\end{split}
\label{eq:dfm_induced}
\end{align}
using the cross product $\mathbf{a}\times\mathbf{b}=e_{ijk}a_{j}b_{k}$
and inner product $\mathbf{a}\cdot\mathbf{b}=a_{i}b_{i}$. 
The first term is the material change produced by the indentation process; 
the second term is an induced increment produced 
by any \emph{tilting increment} $d\mathbf{n}^{m}$
of the particle pair; and the final term is an induced \emph{twirling
increment} which accompanies any rigid rotation of the two particles,
$d\boldsymbol{\theta}^{i}$ and $d\boldsymbol{\theta}^{j}$, about
their common normal axis $\mathbf{n}^{m}$. 
The source density $d\widehat{\mathbf{f}}(\mathbf{n})/dt$
can likewise be viewed as the sum of a material rate and an induced
rotation. 
This analysis is made easier by first treating force density
as the sum of two parts: a normal force density and a tangential force
density,
\begin{equation}
\widehat{\mathbf{f}}(\mathbf{n})=-\widehat{f^{\textrm{n}}}(\mathbf{n})\,\mathbf{n}+\widehat{f^{\textrm{t}}}(\mathbf{n})\,\mathbf{t}(\mathbf{n})\label{eq:fn_plus_ft}
\end{equation}
where $\widehat{f^{\textrm{n}}}(\mathbf{n})$ and $\widehat{f^{\textrm{t}}}(\mathbf{n})$
are scalar densities on the unit sphere, and $\mathbf{t}(\mathbf{n})$
is the average unit direction of tangential force at the particular
orientation $\mathbf{n}$. A compressive normal force $\widehat{f^{\textrm{n}}}$
is considered positive. With this approach, the source density in
Eq.~(\ref{eq:f_rate}) is the sum of two rates,
\begin{equation}
\frac{d\widehat{\mathbf{f}}(\mathbf{n})}{dt}
= \frac{d\widehat{\mathbf{f}^{\textrm{n}}}(\mathbf{n})}{dt}
  +\frac{d\widehat{\mathbf{f}^{\textrm{t}}}(\mathbf{n})}{dt}\label{eq:df_normal_tangent}
\end{equation}
and each vector rate is the sum of a material rate and an induced
rotation, as in Eq.~(\ref{eq:dfm_induced}),
\begin{align}
\begin{split}
\frac{d\widehat{\mathbf{f}^{\textrm{n}}}(\mathbf{n})}{dt}= & -\left(\frac{\partial\widehat{f^{\textrm{n}}}(\mathbf{n})}{\partial t}\right)_{\textrm{matl}}\!\!\!\mathbf{n}
\\
 & -\widehat{f^{\textrm{n}}}(\mathbf{n})\,\mathbf{n}\times\left(\dot{\mathbf{n}}(\mathbf{n})\times\mathbf{n}\right)
\end{split}
\label{eq:fn_source_rate}
\end{align}
\begin{multline}
\frac{d\widehat{\mathbf{f}^{\textrm{t}}}(\mathbf{n})}{dt}=\left(\frac{\partial\widehat{f^{\textrm{t}}}(\mathbf{n})}{\partial t}\right)_{\textrm{matl}}\!\!\!\!\!\mathbf{s}(\mathbf{n})\\
+\widehat{f^{\textrm{t}}}(\mathbf{n})\left[\mathbf{t}(\mathbf{n})\times\left(\dot{\mathbf{n}}(\mathbf{n})\times\mathbf{n}\right)\right]+\left(\frac{d\widehat{\mathbf{f}^{\text{t}}}\left(\mathbf{n}\right)}{dt}\right)_{\text{twirl}}
\label{eq:ft_source_rate}
\end{multline}
\noindent%
The vector field $\mathbf{s}(\mathbf{n})$ is the unit direction
of the tangential material rate, a direction that might differ from
the current direction $\mathbf{t}(\mathbf{n})$ of tangential force
$\widehat{\mathbf{f}^{\text{t}}}(\mathbf{n})$. The final term in
Eq.~\eqref{eq:ft_source_rate} is the net effect of the contacts'
twirling upon the force density. Twirling takes place within the tangent
plane and does not alter the normal force density, so this effect
is absent in Eq.~(\ref{eq:fn_source_rate}).
\par
We now consider the second terms on the right of 
Eqs.~(\ref{eq:f_rate}) and~(\ref{eq:g_rate}), which
include the gradient $\boldsymbol{\nabla}_{\!_{\Omega}}$ 
applied on the curved surface of the unit sphere. These terms account
for divergence and convection effects. 
For example, the gradient of
contact density $\widehat{g}(\mathbf{n})$ in Eq.~(\ref{eq:g_rate}),
as introduced by \citet{Ma:2006a},
can be expanded as two sub-terms: $\boldsymbol{\nabla}_{\!_{\Omega}}\cdot(\dot{\mathbf{n}}\,\widehat{g}(\mathbf{n}))=(\boldsymbol{\nabla}_{\!_{\Omega}}\cdot\dot{\mathbf{n}})\,\widehat{g}(\mathbf{n})+\dot{\mathbf{n}}\cdot\left(\boldsymbol{\nabla}_{\!_{\Omega}}\widehat{g}(\mathbf{n})\right)$.
The first of these sub-terms, the divergence rate $(\boldsymbol{\nabla}_{\!_{\Omega}}\cdot\dot{\mathbf{n}})\,\widehat{g}(\mathbf{n})=\dot{n}_{i,i}\widehat{g}(\mathbf{n})$,
accounts for either a spreading or converging migration that will
rarefy or accumulate contacts at particular orientations $\mathbf{n}$.
In Fig.~\ref{fig:migration}, for example, migration spreads (diverges)
from the compressive $\mathbf{\mathbf{e}}_{1}$ direction and converges
into the extensional $\mathbf{\mathbf{e}}_{3}$ direction. The other
sub-term, the convective rate $\dot{\mathbf{n}}\cdot\boldsymbol{\nabla}_{\!_{\Omega}}\widehat{g}(\mathbf{n})=\dot{n}_{i}\widehat{g}_{,i}$,
addresses the drift of contact density: for example, the convection
of a higher contact density, moving at rate $\dot{\mathbf{n}}$,
toward a lower density, thereby increasing contact density at the
latter orientation.
\par
The gradient of force density that appears in Eq.~(\ref{eq:f_rate})
is similar to the gradient of contact density in Eq.~(\ref{eq:g_rate}),
except that force density is a vector field, whose gradient 
can be analyzed by separating force into its normal
and tangential parts, as in Eq.~(\ref{eq:fn_plus_ft}),
\begin{equation}
\begin{split}
\boldsymbol{\nabla}_{\!_{\Omega}}\cdot\left(\dot{\mathbf{n}}\otimes\widehat{\mathbf{f}}(\mathbf{n})\right)= & -\dot{n}_{j,j}\widehat{f^{\textrm{n}}}n_{i}-\dot{n}_{j}\widehat{f^{\textrm{n}}}_{,j}n_{i}-\dot{n}_{j}\widehat{f^{\textrm{n}}}n_{i,j}\\
 & +\dot{n}_{j,j}\widehat{f^{\textrm{t}}}t_{i}+\dot{n}_{j}\widehat{f^{\textrm{t}}}_{,j}t_{i}+\dot{n}_{j}\widehat{f^{\textrm{t}}}t_{i,j}
\end{split}
\label{eq:grad_force}
\end{equation}
The third and sixth terms can be written as
\begin{align}
-\dot{n}_{j}\widehat{f^{\textrm{n}}}\, n_{i,j}= & -\widehat{f^{\textrm{n}}}(\mathbf{n})\,\dot{\mathbf{n}}(\mathbf{n})
\label{eq:grad_fn_simple}
\\
\dot{n}_{j}\widehat{f^{\textrm{t}}}\, t_{i,j}= & \widehat{f^{\textrm{t}}}(\mathbf{n})\,\dot{\mathbf{t}}(\mathbf{n})
\label{eq:grad_ft_simple}
\end{align}
Because $\dot{\mathbf{n}}$ is normal to $\mathbf{n}$, the 
quantity $\widehat{f^{\textrm{n}}}(\mathbf{n})\,\dot{\mathbf{n}}$
represents a tangential rotation of normal force along a migration
path $\dot{\mathbf{n}}$. The vector field $\dot{\mathbf{t}}(\mathbf{n})$
in Eq.~(\ref{eq:grad_force}) is the change in direction of the tangential
force density along contact migration paths. This vector rate is orthogonal
to $\mathbf{t}(\mathbf{n})$, so that $\dot{\mathbf{t}}(\mathbf{n})$
will have a normal component and, possibly, a tangential component.
A tangential component of $\dot{\mathbf{t}}(\mathbf{n})$ occurs at any
orientation where $\mathbf{t}$ veers in direction along a path $\dot{\mathbf{n}}$:
in Fig.~\ref{fig:migration}, for example, migration arrows are seen
to veer toward the east as they approach the 
$\mathbf{e}_{2}$--$\mathbf{e}_{3}$ equator. 
The two parts of $\widehat{f^{\textrm{t}}}(\mathbf{n})\dot{\mathbf{t}}(\mathbf{n})$,
normal and veering, can be written as
\begin{equation}
\widehat{f^{\textrm{t}}}(\mathbf{n})\,\dot{\mathbf{t}}(\mathbf{n})=-\widehat{f^{\text{t}}}(\mathbf{n})\,\left[\mathbf{t}\cdot\dot{\mathbf{n}}(\mathbf{n})\right]\mathbf{n}+\left(\frac{d\widehat{\mathbf{f}^{\text{t}}}\left(\mathbf{n}\right)}{dt}\right)_{\text{veer}}
\label{eq:ft_w_veer}
\end{equation}
The final, veering part lies in the tangent plane but is orthogonal to
the tangential direction $\mathbf{t}(\mathbf{n})$, as will be demonstrated
in an analysis of DEM data in Section~\ref{sub:Twirl_veer}.
\par
The full rate of force density that produces the stress rate in Eq.~(\ref{eq:stress_rate1})
can be separated into normal and tangential parts,
\begin{equation}
\left.\frac{\partial\widehat{\mathbf{f}}(\mathbf{n})}{\partial t}\right|_{\mathbf{n}}=\left.\frac{\partial\widehat{\mathbf{f}^{\textrm{n}}}(\mathbf{n})}{\partial t}\right|_{\mathbf{n}}+\left.\frac{\partial\widehat{\mathbf{f}^{\textrm{t}}}(\mathbf{n})}{\partial t}\right|_{\mathbf{n}}
\label{eq:ftfn_rate}
\end{equation}
The two parts are expanded by combining Eqs.~(\ref{eq:f_rate}) 
and~(\ref{eq:df_normal_tangent})--\eqref{eq:ft_w_veer}:
\begin{multline}
\left.\frac{\partial\widehat{\mathbf{f}^{\textrm{n}}}(\mathbf{n})}{\partial t}\right|_{\mathbf{n}}=-\left[\left(\frac{\partial\widehat{f^{\textrm{n}}}(\mathbf{n})}{\partial t}\right)_{\textrm{matl}}\!\!\!-\widehat{f^{\textrm{n}}}(\mathbf{n})\left(\boldsymbol{\nabla}_{\!_{\Omega}}\cdot\dot{\mathbf{n}}(\mathbf{n})\right)\right.\\
\left.-\dot{\mathbf{n}}(\mathbf{n})\cdot\left(\boldsymbol{\nabla}_{\!_{\Omega}}\widehat{f^{\textrm{n}}}(\mathbf{n})\right)+\left(\frac{\partial\widehat{f^{\textrm{n}}}(\mathbf{n})}{\partial t}\right)_{\mathrm{diff}}\right]\mathbf{n}
\label{eq:rate_fn}
\end{multline}
and
\begin{align}
\begin{split}
\left.\frac{\partial\widehat{\mathbf{f}^{\textrm{t}}}(\mathbf{n})}{\partial t}\right|_{\mathbf{n}}= & \left(\frac{\partial\widehat{f^{\textrm{t}}}(\mathbf{n})}{\partial t}\right)_{\textrm{matl}}\!\!\mathbf{s}(\mathbf{n})\\
 & -\widehat{f^{\textrm{t}}}(\mathbf{n})\left[\boldsymbol{\nabla}_{\!_{\Omega}}\cdot\dot{\mathbf{n}}(\mathbf{n})\right]\mathbf{t}(\mathbf{n})\\
 & -\dot{\mathbf{n}}(\mathbf{n})\cdot\left(\boldsymbol{\nabla}_{\!_{\Omega}}\widehat{f^{\textrm{t}}}(\mathbf{n})\right)\,\mathbf{t(\mathbf{n})}+\left(\frac{d\widehat{\mathbf{f}^{\text{t}}}\left(\mathbf{n}\right)}{dt}\right)_{\text{twirl}}\\
 & -\left(\frac{d\widehat{\mathbf{f}^{\text{t}}}\left(\mathbf{n}\right)}{dt}\right)_{\text{veer}}+\left(\frac{\partial\widehat{f^{\textrm{t}}}(\mathbf{n})}{\partial t}\right)_{\mathrm{diff}}\!\!\mathbf{t}(\mathbf{n})
\end{split}
\label{eq:rate_ft}
\end{align}
in which we have applied the identity $\mathbf{a}\times(\mathbf{b}\times\mathbf{c})=\mathbf{b}(\mathbf{a}\cdot\mathbf{c})-\mathbf{c}(\mathbf{a}\cdot\mathbf{b})$,
noting that $\mathbf{t}\cdot\mathbf{n}=0$, $\dot{\mathbf{n}}\cdot\mathbf{n}=0$,
and $\mathbf{n}\cdot\mathbf{n}=1$. 
The forms of the material force
rates $\left(\bullet\right)_{\mathrm{matl}}$ and of the twirling
and veering rates are settled by analyzing DEM results in 
Sections~\ref{section:normal_force} and~\ref{section:tangent_force}.
Equations~(\ref{eq:rate_fn}) and~(\ref{eq:rate_ft})
differ from those of \citet{Ma:2006a}, with the inclusion
of the twirling, veering, and diffusions terms and the mutual cancelling
of the tilting terms that arise in 
Eqs.~(\ref{eq:df_normal_tangent})--\eqref{eq:ft_w_veer}.
\par
Until now, we have considered the average contact migration $\dot{\mathbf{n}}$
and its effect on fabric and stress rates. 
We must also consider the random fluctuations
among individual contact motions~--- fluctuations that produce the
diffusion terms in Eqs.~(\ref{eq:g_rate}), (\ref{eq:rate_fn}),
and (\ref{eq:rate_ft}). The tangential rate $\dot{\mathbf{n}}^{m}$
of an individual contact $m$, oriented in the direction $\mathbf{n}^{m}$,
is the sum of the prevailing (mean) migration $\dot{\mathbf{n}}$ and
the individual's fluctuation $\delta\dot{\mathbf{n}}^{m}$ from the mean:
\begin{equation}
\dot{\mathbf{n}}^{m}=\dot{\mathbf{n}}(\mathbf{n}^{m})+\delta\dot{\mathbf{n}}^{m}
\label{eq:Fluctuations}
\end{equation}
\noindent These fluctuations are quite large and can produce a diffusion
of the contact density $\widehat{g}$ during bulk deformation.
Contact diffusion at an orientation $\mathbf{n}$~--- the final term
in Eq.~(\ref{eq:g_rate})~--- is driven by the ongoing deformation
of the bulk material, causing contacts to diffuse (disperse) from
orientations of high contact concentration toward orientations of
lower concentration. This phenomenon is distinct from convection and
divergence~--- the {}``$\boldsymbol{\nabla}_{\!_{\Omega}}$'' term
in Eq.~(\ref{eq:g_rate})~--- in which contacts are swept along
by the prevailing rate $\dot{\mathbf{n}}$. Contact diffusion
can be modelled with the classical diffusion equation, 
in the form
\begin{equation}
\left(\frac{\partial\widehat{g}(\mathbf{n},t)}{\partial t}\right)_{\text{diff}}=D_{g}\nabla^{2}\widehat{g}(\mathbf{n},t){\,\dot{\epsilon}}_{\text{oct}}
\label{eq:ContactDiffusion}
\end{equation}
where $\nabla^{2}\widehat{g}=\widehat{g}_{,kk}$ is the Laplacian
of $\widehat{g}$ on the surface of the unit sphere and $D_{g}$ is
the diffusion coefficient which quantifies the process. 
Because bulk deformation drives the diffusion, 
we use the instantaneous octahedral
strain rate $\dot{\epsilon}_{\text{oct}}$ as a scalar measure of
the distortional strain ($\dot{\epsilon}_{\text{oct}}=\sqrt{D_{ij}^{\prime}D_{ij}^{\prime}/3}$;
$D_{ij}^{\prime}=D_{ij}-\frac{1}{3}D_{kk}\delta_{ij}$), although
other measures could be used as well. In Section~\ref{section:Diffusion_rates},
we provide further reasoning for the form (\ref{eq:ContactDiffusion})
and present the means of extracting the diffusion coefficient $D_{g}$
from DEM simulations.
\par
A diffusion of force density $\widehat{\mathbf{f}}(\mathbf{n})$
will accompany the diffusion of contact density $\widehat{g}(\mathbf{n})$:
contact forces are dispersed with their contacts during deformation. 
The following forms of force diffusion accrue from Eq.~(\ref{eq:avg_f})
and complement the contact diffusion of Eq.~(\ref{eq:ContactDiffusion}):
\begin{equation}
\left(\frac{\partial\widehat{f^{\mathrm{n}}}(\mathbf{n},t)}{\partial t}\right)_{\text{diff}}=D_{g}\overline{f^{\text{n}}}\left(\mathbf{n},t\right)\nabla^{2}\widehat{g}(\mathbf{n},t)\,\dot{\epsilon}_{\mathrm{oct}}
\label{eq:fn_diffusion}
\end{equation}
\begin{equation}
\left(\frac{\partial\widehat{f^{\mathrm{t}}}(\mathbf{n},t)}{\partial t}\right)_{\text{diff}}=D_{g}\overline{f^{\text{t}}}\left(\mathbf{n},t\right)\nabla^{2}\widehat{g}(\mathbf{n},t)\,\dot{\epsilon}_{\mathrm{oct}}
\label{eq:ft_diffusion}
\end{equation}
\noindent which define the scalar diffusion rates that appear in
Eqs.~(\ref{eq:rate_fn}) and~(\ref{eq:rate_ft}).
\par
To summarize this section, 
the rates of contact and force densities include material
rates, combined with divergence, twirling, 
and diffusion effects (Eqs.~\ref{eq:f_rate} and~\ref{eq:g_rate}). 
As will be seen, a sort of competition exists
between the material rate and the other effects. Anisotropy is usually
reduced by divergence, twirling, and diffusion, 
as these effects tend to diminish
$\widehat{g}$ and $\widehat{\mathbf{f}}$ at those orientations $\mathbf{n}$
where the densities are large. At the same orientations, the densities
are usually replenished by the material rates, which represent a generation
of contacts and of contact force. The material rate will be seen to
dominate at the start of loading, inducing fabric anisotropy and deviatoric
stress. During failure, the material rate becomes 
quite small and is counteracted
by the other rate effects, eventually leading to a steady, critical
state of fabric and stress.
\section{\label{sec:Quantifying}Quantifying fabric evolution}
In the previous section, we found that the stress rate $\dot{\boldsymbol{\sigma}}$
results from changes in the contact force density
$\widehat{\mathbf{f}}(\mathbf{n})$~--- changes 
that can be separated into several rate fields. 
In this section, specific forms of the various fields are adopted, 
guided by DEM observations of granular behavior. 
Taken together, these forms comprise the rudiments
of a constitutive model for soils and other granular materials, a
model based on micro-mechanics and informed by DEM results. 
The Paper focuses on behavior at the critical state, 
when granular materials
reach a stationary condition: during sustained steady state flow, 
the volume, the stress, the fabric, and, most notably, 
the density functions $\widehat{g}(\mathbf{n})$,
$\widehat{f^{\text{n}}}(\mathbf{n})$, 
and $\widehat{\mathbf{f}^{\text{t}}}(\mathbf{n})$
remain constant. 
To aid understanding of the critical state behavior,
we also consider the rate of $\widehat{f^{\text{n}}}(\mathbf{n})$
at the other extreme of deformation~--- at the start of loading~---
which will provide a basis for quantifying 
the corresponding critical state terms. 
We begin with a brief description of the simulations.
\subsection{\label{section:DEM}DEM simulations}
\begin{figure}
\centering
\includegraphics{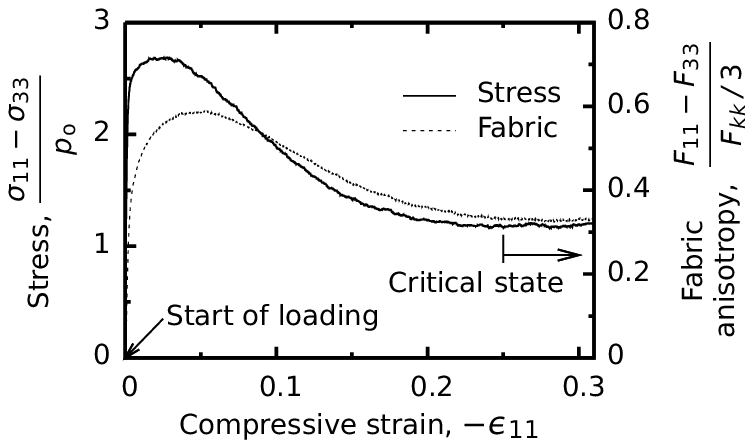}
\caption{Average behavior of twenty DEM assemblies in biaxial
plane-strain compression (see inset in Fig.~\ref{fig:migration}).}
\label{fig:crs}
\end{figure}

DEM simulations were performed on twenty small assemblies of spherical
particles. 
The simulations permitted the observation of a sufficiently large
number of contacts to attain their motions, force rates,
and net creation rates across the entire unit sphere of orientations.
The assemblies contained the same set of 4096 particles
that were randomly packed into cube containers having periodic boundaries
on all sides and the dense initial conditions listed in Table~\ref{table:DEM}.%
\begin{table*}
\centering
\begin{tabular}{lc}
\hline 
Characteristic & Value\\
\hline
Assemblies & 20\\
Assembly shape & cube\\
Assembly particles & 4096\\
Assembly dimension & 13.4$D_{50}$\\
Assembly boundaries & periodic\\
Particle shape & spherical\\
Particle size range & 0.4$D_{50}$ -- 1.2$D_{50}$\\
Particle shear modulus, $G$ & 29~GPa\\
Particle Poisson ratio, $\nu$ & 0.15\\
Inter-particle friction ratio & 0.50\\
Initial particle arrangement & dense, isotropic\\
Initial void ratio, solids fraction & 0.510, 0.662\\
Initial avg. pressure, $p_{\text{o}}=-\sigma_{ii}/3$ & 320~kPa\\
Initial avg. coord. no., $2M/N$ & 5.48\\
\hline
\end{tabular}
\caption{Characteristics of DEM assemblies}
\label{table:DEM}
\end{table*}
The spheres were polydisperse with diameters ranging from 0.4$D{}_{50}$
to 1.5$D{}_{50}$, where $D{}_{50}$ is the median diameter. The hard
non-breaking particles interact at Hertz contacts having a frictional
limit $\mu=0.5$ and a modified Mindlin tangential stiffness, as described
by \citet{Lin:1997a}. The cubic assemblies had dimensions of about
$13.4\times13.4\times13.4$ particle diameters: small enough to prevent
shear bands, yet large enough to capture the average, bulk material behavior.
Because a small assembly of only 4096 particles will exhibit substantial
spikes in stress during deviatoric loading, twenty different assemblies
were randomly created and then loaded. Their averaged behavior is
reported herein. Although several loading paths are studied, the primary
loading was slow biaxial plane-strain compression: the assemblies
were compressed by continually reducing their $x_{1}$ dimension at
a constant rate ($\dot{\epsilon}_{11}=\textrm{constant}$) while
maintaining a constant normal stress in the $x_{3}$ direction and
a constant assembly width in the $x_{2}$ direction ($\sigma_{33}$
constant $=p_{\text{o}}=320\:\text{kPa}$, and $\epsilon_{22}=0$;
see inset in Fig.~\ref{fig:migration}). Because expansion was permitted
against a constant stress $\sigma_{33}$, the simulations can be considered
as drained tests in the geotechnical sense. 
The averaged results are displayed in Fig.~\ref{fig:crs},
which shows the normalized deviator stress $(\sigma_{11}-\sigma_{33})/p_{\text{o}}$
and the fabric anisotropy $F_{11}-F_{33}$ during the course of plane-strain
biaxial compression (the fabric measure $F_{ij}$ 
is defined as 
$\int_{\Omega}n_{i}n_{j}\widehat{g}(\mathbf{\mathbf{n})\,}d\Omega$,
as in \citealt{Nemat-Nasser:2001b}).%
Because we are interested in the rates of fabric and stress evolution
at both the macro (bulk) and micro scales, contact statistics were
compiled during loading at the critical state (at $-\epsilon_{11}>0.25)$.
{}``Snapshots'' of the contacts and their rates were taken at several
such strains, and by doing the same for all twenty assemblies, we were
able to analyze over 800,000 contacts, as discussed below.
\subsection{\label{section:Migration_rate}Migration rate $\dot{\mathbf{n}}$,
convection, and divergence}
During deviatoric loading, particles roll and slide across each other,
causing contact orientations to migrate. DEM simulations were used
to determine a functional form of the average
migration $\dot{\mathbf{n}}(\mathbf{n})$
in relation to the bulk deformation rate. 
As an example, the migrations
of 800,000 contacts were measured at the critical state in simulations
of plane-strain biaxial compression, and the average rates are depicted
in Fig.~\ref{fig:migration}. Although somewhat obscured, two arrows
emanate from each grid point. The heavier arrows are the actual, measured
rates. These arrows are closely aligned with lighter arrows that represent
a certain projection $\dot{\mathbf{n}}^{\mathrm{proj}}$ of the instantaneous
deformation rate $\mathbf{D}$ onto the unit sphere:
\begin{align}
\begin{split}
\dot{\mathbf{n}}(\mathbf{n})\:\overset{\mathrm{\stackrel{{\scriptstyle aligned}}{with}}}{\longleftrightarrow}\:\dot{\mathbf{n}}^{\mathrm{proj}}(\mathbf{n}) & \equiv\mathbf{P}^{\text{n}}(\mathbf{n})\cdot(\mathbf{D}\cdot\mathbf{n})
\\
 & =(\mathbf{I}-\mathbf{n}\otimes\mathbf{n})\cdot(\mathbf{D}\cdot\mathbf{n})
\end{split}
\label{eq:n_align}
\end{align}
with $\mathbf{P}^{\text{n}}\cdot\mathbf{D}\cdot\mathbf{n}=P_{ij}^{\text{n}}D_{jk}n_{k}$.
In this equation, $\mathbf{D}$ is the rate of deformation tensor
(instantaneous strain rate), and the matrix $\mathbf{P}^{\text{n}}(\mathbf{n})=\delta_{ij}-n_{i}n_{j}$
projects the rate vector $\mathbf{D}\cdot\mathbf{n}$ ($=D_{jk}n_{k}$)
onto the tangent plane. The rate $\dot{\mathbf{n}}^{\mathrm{proj}}$
represents the ideal tangential motion of two contacting spheres whose
centers move in perfect accord with the mean, bulk deformation $\mathbf{D}$.
Although individual contacts migrate in widely varying directions
and rates, the alignment of the average rate $\dot{\mathbf{n}}$ and
the ideal rate $\dot{\mathbf{n}}^{\mathrm{proj}}$ is quite close:
the two directions differ, on average, by less than $3^{\circ}$.
\par
Although the observed and ideal rates are aligned, they differ in
magnitude. Different scales have been applied in Fig.~\ref{fig:migration}
for displaying the measured rates $\dot{\mathbf{n}}$ and the projected
rates $\dot{\mathbf{n}}^{\mathrm{proj}}$, and although obscured in
the small figure, the lengths of $\dot{\mathbf{n}}$ and $\dot{\mathbf{n}}^{\mathrm{proj}}$
are consistently in about the same proportion: the two vector fields
are nearly aligned and proportional. This observation suggests that
the migration rate $\dot{\mathbf{n}}$ can be approximated as
\begin{equation}
\dot{\mathbf{n}}\approx\alpha\,\dot{\mathbf{n}}^{\mathrm{proj}}
\label{eq:ndot_alpha}
\end{equation}
a condition that is closely held throughout the loading process. At
the start of loading, the simulations show that the factor $\alpha$
is equal to 1.0, so that the measured and ideal rates are about equal
(the assumption of $\dot{\mathbf{n}}=\dot{\mathbf{n}}^{\mathrm{proj}}$
being made by \citealt{Didwania:2001a}).
As deformation progresses to larger strains, 
the actual rate exceeds the ideal projected
rate, with $\alpha=1.7$ at the critical state.
\par
The approximation of Eq.~(\ref{eq:ndot_alpha}) 
can be used to estimate the convection and divergence
rates in Eqs.~(\ref{eq:g_rate}), (\ref{eq:rate_fn}), and~(\ref{eq:rate_ft}).
For example, the scalar divergence of $\dot{\mathbf{n}}$ is
\begin{equation}
\boldsymbol{\nabla}_{\!_{\Omega}}\cdot\dot{\mathbf{n}}\approx\boldsymbol{\nabla}_{\!_{\Omega}}\cdot(\alpha\dot{\mathbf{n}}^{\mathrm{proj}})=-3\alpha\mathbf{n}\cdot(\mathbf{D}^{\prime}\cdot\mathbf{n})\label{eq:divergence_ndot}
\end{equation}
or $-3\alpha D_{ij}^{\prime}n_{i}n_{j}$, where $\mathbf{D}^{\prime}$
($=D_{ij}-D_{kk}\delta_{ij}/3$) is the deviatoric part of deformation.
\subsection{\label{section:normal_force}Rate of normal force}
Particles press against each other with changing force
as a granular material undergoes bulk deformation. The material rate
of normal force density, $(\partial\widehat{f^{\textrm{n}}}(\mathbf{n})/\partial t)_{\textrm{matl}}$,
is the net effect of these many changes at particular orientations
$\mathbf{n}$. To quantify this material rate, several hundreds of
thousands of contacts were observed in DEM simulations of biaxial
plane-strain loading at two extremes of deformation: at the start
of loading and during sustained flow at the critical state. In general,
the material rate exhibits the following characteristic: the rate
is usually positive (increasingly compressive) at orientations $\mathbf{n}$
in which the bulk strain produces compression; whereas, the material
rate is negative (tensile) in directions of extension. This observation,
although not surprising, 
suggests that the \emph{average} ``$\,\overline{\rule{0em}{1ex}\;\;}\,$'' compressive rate,
$(\partial\overline{f^{\textrm{n}}}(\mathbf{n})/\partial t)_{\textrm{matl}}$,
might be approximated as the product of an average normal stiffness
$\overline{k^{\textrm{n}}}$ and the average rate of approach between
the centers of contacting particles:
\begin{equation}
\left(\frac{\partial\overline{f^{\textrm{n}}}(\mathbf{n})}{\partial t}\right)_{\text{matl}}=-\overline{k^{\text{n}}}(\mathbf{n})\,\frac{\partial\ell(\mathbf{n})}{\partial t}
\label{eq:dfn_avg}
\end{equation}
If the particle motions were to conform to the mean, bulk rate of
deformation field $\mathbf{D}$, 
the rate of approach would be $\partial\ell(\mathbf{n})/\partial t=\overline{\ell}(\mathbf{n})\,\mathbf{n}\cdot(\mathbf{D}\cdot\mathbf{n})=\overline{\ell}(\mathbf{n})\, n_{i}D_{ij}n_{j}$,
where $\overline{\ell}(\mathbf{n})$ is the average distance between
the centers of particles oriented in direction $\mathbf{n}$. 
Numerous studies have sought expressions for the bulk elastic moduli
of granular media by starting with this mean-field assumption.
This approach over-estimates the moduli and is usually amended by 
considering motion fluctuations from the mean
(e.g.~\citealt{Jenkins:2005a}).
Successful estimates are only achieved, however, at small strains and while
behavior is elastic.
Observations have shown that the normal motions between particles,
$\partial\ell(\mathbf{n})/\partial t$,
are suppressed during deformation~--- especially at large strain~---
as particles tend to roll and slide in a manner that minimizes such
motion \citep{Kuhn:2004k}. In this regard, we will introduce a 
factor $\beta^{\text{n}}$ to reduce the indentation rate between particles.
With this change, the scalar rate $\partial\ell(\mathbf{n})/\partial t$
is approximated as
\begin{equation}
\frac{\partial\ell(\mathbf{n})}{\partial t}\approx\beta^{\text{n}}\,\overline{\ell}\,\mathbf{n}\cdot(\mathbf{D}\cdot\mathbf{n})
\label{eq:dl_dt}
\end{equation}
In this approximation, we also use the average branch length $\overline{\ell}$
in place of the function $\overline{\ell}(\mathbf{n})$, ignoring
the small correlation between branch length and orientation.
\par
The average contact stiffness $\overline{k^{\text{n}}}$ in Eq.~(\ref{eq:dfn_avg})
will depend upon the stiffness characteristics of the particles themselves.
For a single $m$th contact between two isotropic elastic spheres
of equal size, the Hertz stiffness is
\begin{equation}
k^{\textrm{n},m}=\left[\frac{3G^{2}R^{m}f^{\textrm{n},m}}{(1-\nu)^{2}}\right]^{1/3}
\label{eq:k_nm}
\end{equation}
where $f^{\textrm{n},m}$ is the pair's current normal (compressive)
force, $G$ is the particle shear modulus, $\nu$ is the Poisson's ratio, and
$R^{m}$ is the shared radius. This contact stiffness depends upon
the contact force $f^{\textrm{n},m}$. Because the average normal
force $\overline{f^{\textrm{n}}}(\mathbf{n})$ within an entire assembly
is known to be anisotropic, we would expect the average contact
stiffness $\overline{k^{\text{n}}}$ to depend on orientation. The
average stiffness among all contacts that share an orientation $\mathbf{n}$
can be approximated as
\begin{equation}
\overline{k^{\textrm{n}}}(\mathbf{n})\approx\left[\frac{3G^{2}\overline{\ell}\,\overline{f^{\textrm{n}}}(\mathbf{n})}{2(1-\nu)^{2}}\right]^{1/3}
\label{eq:kn}
\end{equation}
using $\overline{\ell}/2$ in place of $R$. By combining Eqs.~(\ref{eq:dfn_avg}),
(\ref{eq:dl_dt}), and~(\ref{eq:kn}) and twice applying Eq.~(\ref{eq:avg_f}),
we can estimate the material rate of normal force density as
\begin{equation}
\left(\frac{\partial\widehat{f^{\text{n}}}(\mathbf{n})}{\partial t}\right)_{\text{matl}}
\approx
-\beta^{\text{n}}\,\overline{\ell}\,\widehat{k^{\text{n}}}(\mathbf{n})\left[\mathbf{n}\cdot(\mathbf{D}\cdot\mathbf{n})\right]
\label{eq:fn_rate_matl}
\end{equation}
with the \emph{stiffness density}
\begin{equation}
\widehat{k^{\text{n}}}(\mathbf{n})=\overline{k^{\textrm{n}}}(\mathbf{n})\,\widehat{g}(\mathbf{n})=\left[\frac{3G^{2}\overline{\ell}}{2(1-\nu)^{2}}\,\widehat{f^{\text{n}}}(\mathbf{n})\left(\widehat{g}(\mathbf{n})\right)^{2}\right]^{1/3}
\label{eq:kn_hat}
\end{equation}
Coefficient $\beta^{\text{n}}$ was investigated for two extremes
of loading: during initial loading and during sustained plastic flow
at the critical state.
\subsubsection{Rate of normal force at small strains}
\begin{figure}
\centering
\includegraphics{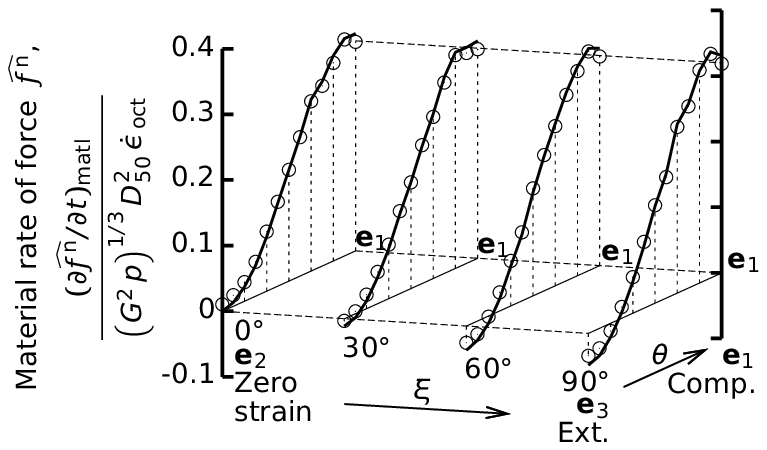}
\caption{Material rates of normal force creation for
plane-strain biaxial compression at the start of loading: data from
DEM simulations (symbols~$\circ$), and Eq.~(\ref{eq:fn_rate_matl})
with $\beta^{\text{n}}=0.94$ (lines~---). Data is arranged along
meridians of $30{}^{\circ}$ spacing (see Fig.~\ref{fig:migration}).}
\label{fig:fn_Zero}
\end{figure}

Upon initial loading of dense DEM sphere packings, 
the value $\beta^{\text{n}}$
was measured as 0.94: the actual stiffness was slightly smaller than
the ideal stiffness that would apply if each particle moved
in perfect accord with the mean deformation field. Figure~\ref{fig:fn_Zero}
compares Eq.~(\ref{eq:fn_rate_matl}) and the DEM data, with both
plotted in a dimensionless form.%
Four meridians of the unit sphere are shown in the figure, corresponding
to the angle $\xi$ in Fig.~\ref{fig:migration}. 
Equation~(\ref{eq:fn_rate_matl}) closely fits the DEM data.
\par
With a $\beta^{\text{n}}=0.94$ at the start of loading, the material
rate of normal force density is more than two hundred times larger
than the other rates that contribute to the force density: the divergence,
convection, and diffusion terms in Eq.~(\ref{eq:rate_fn}). The rapid
evolutions of fabric and stress at the start of loading are, therefore,
dominated by the material rate $(\partial\widehat{f^{\text{n}}}(\mathbf{n})/\partial t)_{\textrm{matl}}$,
with the other rates nearly inconsequential. 
The situation changes, however, upon further loading. 
Because $\beta^{\text{n}}$ decreases with increasing
strain, its hardening effect is progressively diminished, 
and the influences of divergence,
convection, and diffusion become increasingly more significant~--- nearly
dominant~--- as will be seen in the next paragraphs.
\subsubsection{\label{sub:Normal_large}Rate of normal force at the critical state}
DEM simulations were also used to 
measure the material parameter $\beta^{\text{n}}$
during failure at the critical state. 
Because the total rate of normal force $\left.\partial\widehat{\mathbf{f}^{\text{n}}}/\partial t\right|_{\mathbf{n}}$
is zero at the critical state, the corresponding material rate can
be readily computed from the remaining terms in Eq.~(\ref{eq:rate_fn}).
A $\beta^{\text{n}}=0.0037$ fits the DEM data, although this value
must be slightly amended, as described below. Such a small value of
$\beta^{\text{n}}$ indicates that the average normal motions between
particles, the rate $\partial\ell(\mathbf{n})/\partial t$ in Eq.~(\ref{eq:dl_dt}),
is much smaller than would be anticipated by assuming that the particle
motions conform to a uniform, affine deformation field. 
On the other hand, 
we had previously noted that the average tangential motions
become somewhat \emph{greater} than those of uniform deformation, 
with an $\alpha$ of 1.7 in Eq.~(\ref{eq:ndot_alpha}). 
These contrasting
results are consistent with other evidence that the large-strain motions
of particles are dominated by the tangential rolling of particle pairs,
but with minimal average normal motions at the contacts \citep{Kuhn:2004k}.
\par
The DEM simulations also reveal an unexpected aspect of the material
rate $(\partial\widehat{f^{\textrm{n}}}(\mathbf{n})/\partial t)_{\textrm{matl}}$:
although one might expect a force rate of zero at those neutral orientations
$\mathbf{n}$ where particles neither approach nor withdraw (where
$\partial\ell(\mathbf{n})/\partial t=0$), we find instead that the
material rate is usually slightly negative 
(depletive or tensile) at these orientations.
This anomalous situation is most noticeable at neutral orientations
that also have large migration rates $\dot{\mathbf{n}}(\mathbf{n})$. To
account for this observation, we apply a small adjustment to the material
rate of Eq.~(\ref{eq:fn_rate_matl}), replacing that equation as
follows:
\begin{align}
\begin{split}
\left(\frac{\partial\widehat{f^{n}}(\mathbf{n})}{\partial t}\right)_{\text{matl}}\approx & -\beta^{\text{n}}\,\widehat{k^{\text{n}}}(\mathbf{n})\,\overline{\ell}\left[\mathbf{n}\cdot(\mathbf{D}\cdot\mathbf{n})\right]
\\
 & -\,\beta_{2}^{\text{n}}\,\widehat{f^{\textrm{n}}}(\mathbf{n})\frac{\left|\dot{\mathbf{n}}(\mathbf{n})\right|^{2}}{\dot{\epsilon}_{\mathrm{oct}}}
\end{split}
\label{eq:fn_rate_matl2}
\end{align}
in which the new, subtracted $\beta_{2}^{\text{n}}$ term produces
a small depletive bias in the material rate. The magnitude $\left|\dot{\mathbf{n}}(\mathbf{n})\right|$
is approximated with Eq.~(\ref{eq:ndot_alpha}) and is normalized
by dividing by the average rate $\dot{\epsilon}_{\mathrm{oct}}=\sqrt{D_{ij}^{\prime}D_{ij}^{\prime}/3}$.
The factor $\beta_{2}^{\text{n}}$ was measured as 0.65.
\par
We must also consider the role of the mean stress in generating contact
force density. For spherical particles, the mean stress depends exclusively
on the normal components of the contact forces \citep{Cundall:1983a},
such that the bulk pressure $p$ is proportional to the average normal
force density $\widehat{f^{\text{n}}}(\mathbf{n})$. At small strains,
the total rate of normal force density, the rate $\partial\widehat{f^{\text{n}}}(\mathbf{n})/\partial t$
on the left of Eq.~(\ref{eq:rate_fn}), is dominated by the material
rate, and its approximation with Eqs.~(\ref{eq:kn_hat}) and~(\ref{eq:fn_rate_matl2})
would suggest that the bulk stiffness is proportional to $G^{2/3}$
and to $p^{1/3}$. This scaling at small strains is in fair agreement
with small-strain vibrational experiments which show that the elastic
moduli are proportional to $p^{n}$ with an exponent $n$ between
1/3 and 1/2 (see \citealt{Goddard:1990a} for a review). Granular
behavior at large strains scales quite differently.
At the critical state, strength is proportional to the confining pressure,
$p^{1}$, and is nearly independent of the particle stiffness $G$. If
left unmodified, the stiffness density in Eq.~(\ref{eq:kn_hat})
would produce a strength proportional to $p^{1/3}$ and to $G^{2/3}$,
contrary to the observed behavior of soils and other granular materials. 
We should expect, therefore,
that at large strains, the factor $\beta^{\text{n}}$ will depend
upon $p$ and upon the particles' elastic properties in the following
manner:
\begin{equation}
\beta^{\text{n}}=B^{\text{n}}\left(\frac{p\,(1-\nu)}{G}\right)^{2/3}
\label{eq:beta_B}
\end{equation}
where the dimensionless factor $B^{\text{n}}$ effects a proper scaling
of granular strength at large strains. We tested this hypothesis by
running DEM simulations of biaxial compression that were identical
to those previously described, except that the initial confining pressure
was increased about six-fold. At the critical state, the mean stress
increased from 490~kPa to 3200~MPa and the strength was found to
increase by the same factor, but $\beta^{\text{n}}$ had only increased
from 0.0037 to 0.012~--- not a six-fold increase, but roughly in
accord with Eq.~(\ref{eq:beta_B}) and a $B^{\text{n}}=6.3$. The
second parameter in Eq.~(\ref{eq:fn_rate_matl2}), $\beta_{2}^{\text{n}}$,
remained about the same for the two confining pressures, as would
be expected, since the last term in Eq.~(\ref{eq:fn_rate_matl2})
is proportional to $\widehat{f^{\text{n}}}$, a form that is consistent
with strength being proportional to mean stress.
\par
\begin{figure}
\begin{centering}
\includegraphics{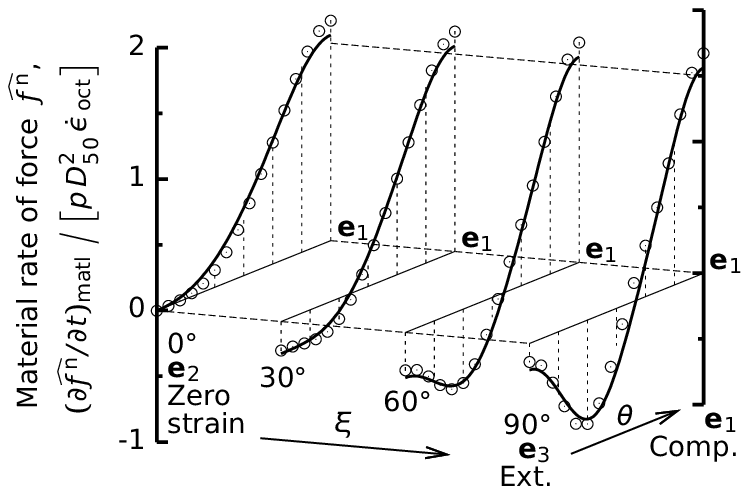}
\par\end{centering}
\caption{Material rates of normal force creation during plane-strain
biaxial compression at the critical state: data from DEM simulations
(symbols~$\circ$), and Eqs.~(\ref{eq:fn_rate_matl2}) and~(\ref{eq:beta_B}),
with $B^{\text{n}}=6.3$ and $\beta_{2}^{\text{n}}=0.65$ (lines~---).
Data is arranged along meridians of $30{}^{\circ}$ 
spacing (see Fig.~\ref{fig:migration}).}
\label{fig:fn_ft_rates}
\end{figure}

In Fig.~\ref{fig:fn_ft_rates}, 
the combination of Eqs.~(\ref{eq:fn_rate_matl2})
and~(\ref{eq:beta_B})
is compared with data from 800,000 contacts in DEM simulations. %
The simulations are of plane-strain biaxial compression at the critical
state, and data is presented along four meridians (angle $\xi$, Fig.~\ref{fig:migration}).
All results are reported in a dimensionless form: force density is
divided by the stress $p$, and its time rate is normalized with respect
to the average octahedral rate $\dot{\epsilon}_{\mathrm{oct}}$. 
Equations~(\ref{eq:fn_rate_matl2}) and~(\ref{eq:beta_B})
are in close agreement with the DEM data.
\subsection{\label{section:tangent_force}Rates of tangential force}
DEM simulations can be used to resolve a realistic form for the material
rate of tangential force at the critical state~--- the rate $(\partial\widehat{f^{\text{t}}}(\mathbf{n})/\partial t)_{\textrm{matl}}$
in Eq.~(\ref{eq:rate_ft}). We will quantify this material rate,
after first considering the tangential force rates of twirling and
veering.
\subsubsection{\label{sub:Twirl_veer}Twirling and veering rates at the critical
state}
\begin{figure}
\centering
\includegraphics{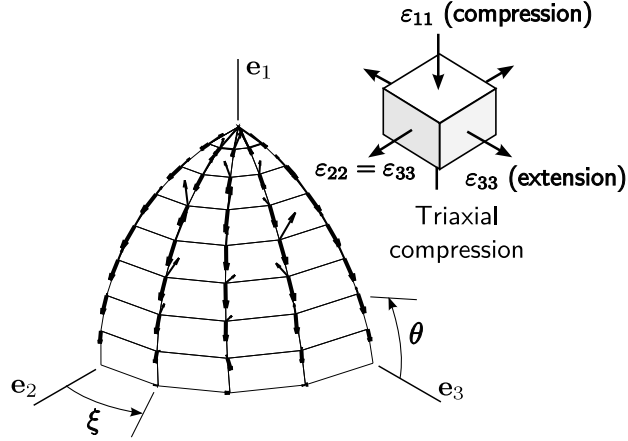}
\caption{DEM simulations of triaxial compression
at the critical state: (1)~the average twirling rates of tangential
contact forces (upward, light arrows), and (2)~the contact migration
rates (downward, darker arrows).}
\label{fig:Twirl_sphere_eps}
\end{figure}

During sustained flow at the critical state, the DEM simulations show
that the tangential force density 
$\widehat{f^{\text{t}}}(\mathbf{n})\mathbf{t}(\mathbf{n})$
becomes closely aligned with the direction of contact migration $\dot{\mathbf{n}}(\mathbf{n})$,
such that the unit direction is approximated as
\begin{equation}
\mathbf{t}(\mathbf{n})\approx\dot{\mathbf{n}}\left/\left|\dot{\mathbf{n}}\right|\right.
\label{eq:t_ndot}
\end{equation}
where $\dot{\mathbf{n}}(\mathbf{n})$ is the field depicted in Fig.~\ref{fig:migration}
and given by Eqs.~(\ref{eq:n_align}) and~(\ref{eq:ndot_alpha}).
The total rate of tangential force that appears on the left of Eq.~(\ref{eq:rate_ft})
results from various effects: a material rate combined with convection,
twirling, veering, and diffusion effects. The density rates of twirling
and veering involve rotations of tangential force within the tangent
plane. 
We can isolate and directly measure the twirling rate by considering other
DEM simulations that eliminate the veering effect: by using triaxial
rather than plane-strain loading. 
With triaxial compression, assemblies
are compressed in the $\mathbf{e}_{1}$ direction while 
constant stress is maintained in the 
$\mathbf{e}_{2}$ and $\mathbf{e}_{3}$ directions
(see inset, Fig.~\ref{fig:Twirl_sphere_eps}).
This symmetric loading condition, in which two principal strains are
equal, produces contact migrations $\dot{\mathbf{n}}$, as defined
in Eq.~(\ref{eq:n_align}), along meridians (geodesics) that emanate
from the $\mathbf{e}_{1}$ pole 
and approach the $\mathbf{e}_{2}$--$\mathbf{e}_{3}$ equator. 
As stated above, $\mathbf{t}(\mathbf{n})$ and $\dot{\mathbf{n}}(\mathbf{n})$
are aligned at the critical state, and because neither direction veers
under triaxial loading, we can use triaxial DEM data to directly measure
any bulk twirling of the tangential forces. 
\par
\begin{figure}
\centering
\includegraphics{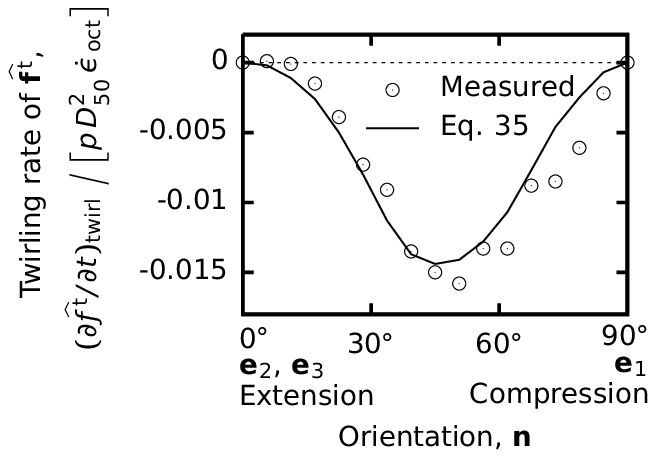}
\caption{Twirling rates of tangential force in triaxial
compression at the critical state: 
comparison of DEM data and Eq.~(\ref{eq:ft_twirl}),
with $\gamma_{\text{twirl}}=1.0$. }
\label{fig:ft_twirl}
\end{figure}

The twirling rates (i.e., from
the final term in Eq.~\ref{eq:dfm_induced}) of over 800,000 contacts
were averaged, and Fig.~\ref{fig:Twirl_sphere_eps} illustrates these
average density rates.%
Although somewhat obscured in the small monochrome figure, two vector
fields are displayed: thinner arrows correspond to the twirling density
$(\partial\widehat{\mathbf{f}^{\text{t}}}(\mathbf{n})/\partial t)_{\textrm{twirl}}$;
thicker arrows are the contact migration field $\dot{\mathbf{n}}(\mathbf{n})$.
Some scatter is apparent in the data, but the results indicate that
the twirling rate vectors are consistently in a direction opposite
the migration field. For example, in the northern octant 
of Fig.~\ref{fig:Twirl_sphere_eps},
tangential forces tend to rotate toward
the north, even as their contacts are migrating toward the south. 
The magnitude of the twirling density was also found to be roughly proportional
to the product of the tangential force density $\widehat{f^{\text{t}}}(\mathbf{n})$
and the magnitude of $\dot{\mathbf{n}}(\mathbf{n})$. 
These observations
suggest the following form of the twirling density field:
\begin{equation}
\left(\frac{\partial\widehat{\mathbf{f}^{\textrm{t}}}(\mathbf{n})}{\partial t}\right)_{\textrm{twirl}}\approx-\gamma_{\text{twirl}}\,\widehat{f^{\text{t}}}(\mathbf{n})\,\dot{\mathbf{n}}(\mathbf{n})
\label{eq:ft_twirl}
\end{equation}
The factor $\gamma_{\text{twirl}}$ was about 1.0 in the DEM simulations.
Figure~\ref{fig:ft_twirl} compares this equation with the twirling
data for triaxial compression at the critical state. %
Equation~(\ref{eq:ft_twirl}) and a $\gamma_{\text{twirl}}=1.0$ 
fit the data, although the figure
also indicates that the twirling rate of force is relatively small
when compared with the normal force rates that are shown in 
Fig.~\ref{fig:fn_ft_rates}.
\par
Having resolved the twirling effect, we can now investigate possible
veering of the tangential force direction $\mathbf{t}(\mathbf{n})$.
To this end, we return to biaxial plane-strain compression simulations.
The veering rate of tangential force is the tangential component of
the rate $\widehat{f^{\text{t}}}\dot{\mathbf{t}}$ along migration
paths $\dot{\mathbf{n}}$ (from Eqs.~\ref{eq:grad_ft_simple}
and~\ref{eq:ft_w_veer}): 
\begin{equation}
\left(\frac{d\widehat{\mathbf{f}^{\text{t}}}\left(\mathbf{n}\right)}{dt}\right)_{\text{veer}}=\widehat{f^{\text{t}}}(\mathbf{n})\,\mathbf{P}^{\text{n}}(\mathbf{n})\cdot\left(\frac{\partial\mathbf{t}(\mathbf{n})}{\partial\mathbf{n}}\cdot\dot{\mathbf{n}}(\mathbf{n})\right)
\notag
\end{equation}
where the projection matrix $\mathbf{P}^{\text{n}}(\mathbf{n})$ extracts
the tangential component of this rate (see Eq.~\ref{eq:n_align}).
As an intermediate step in deriving the veering rate, 
the unit vector $\mathbf{t}(\mathbf{n})$
can be expressed as the product of a matrix $\mathbf{Q}(\mathbf{n})$
and the unit normal $\mathbf{n}$, as 
\mbox{$\mathbf{t}(\mathbf{n})=\mathbf{Q}(\mathbf{n})\cdot\mathbf{n}/|\mathbf{Q}(\mathbf{n})\cdot\mathbf{n}|$},
in which $\mathbf{Q}\cdot\mathbf{n}=Q_{ij}n_{j}$. After differentiating
$\mathbf{t}(\mathbf{n})$ with respect to $\mathbf{n}$,
\begin{multline}
\left(\frac{d\widehat{\mathbf{f}^{\text{t}}}\left(\mathbf{n}\right)}{dt}\right)_{\text{veer}}=\\
\widehat{f^{\text{t}}}(\mathbf{n})\,\mathbf{P}^{\text{n}}(\mathbf{n})\cdot\mathbf{P}^{\text{t}}(\mathbf{n})\cdot\left(\frac{\frac{\partial\mathbf{Q}(\mathbf{n})}{\partial\mathbf{n}}\cdot\mathbf{n}+\mathbf{Q}(\mathbf{n})}{\left|\mathbf{Q}(\mathbf{n})\cdot\mathbf{n}\right|}\cdot\dot{\mathbf{n}}(\mathbf{n})\right)
\label{eq:veer_Q}
\end{multline}
where the new operator $\mathbf{P}^{\text{t}}(\mathbf{n})=\mathbf{I}-\mathbf{t}\otimes\mathbf{t}=\delta_{ij}-t_{i}t_{j}$
projects vectors onto a plane that is perpendicular to the tangent
direction $\mathbf{t}$. Combined with $\mathbf{P}^{\text{n}}(\mathbf{n})$,
the matrix product $\mathbf{P}^{\text{n}}\cdot\mathbf{P}^{\text{t}}=P_{ik}^{\text{n}}P_{kj}^{\text{t}}$
produces a veering rate that is orthogonal to both $\mathbf{n}$ and
$\mathbf{t}$.
\par
As expressed in Eq.~(\ref{eq:t_ndot}), 
the simulations show that during sustained flow
at the critical state, the tangential force density is closely aligned
with the migration direction $\dot{\mathbf{n}}=\mathbf{P}^{\text{n}}\cdot\mathbf{D}\cdot\mathbf{n}$,
defined in Eqs.~(\ref{eq:n_align}) and (\ref{eq:ndot_alpha}): that is,
matrix $\mathbf{Q}(\mathbf{n})$ is equal to the matrix product
$\mathbf{P}^{\text{n}}(\mathbf{n})\cdot\mathbf{D}$.
The corresponding veering rate in Eq.~(\ref{eq:veer_Q})
is, therefore,
\begin{multline}
\left(\frac{d\widehat{\mathbf{f}^{\text{t}}}\left(\mathbf{n}\right)}{dt}\right)_{\text{veer}}=\widehat{f^{\text{t}}}(\mathbf{n})\,\mathbf{P}^{\text{n}}(\mathbf{n})\cdot\mathbf{P}^{\text{t}}(\mathbf{n})\\
\cdot\left(\frac{\mathbf{D}-2\left(\mathbf{n}\otimes\mathbf{n}\right)\cdot\mathbf{D}-\left[\mathbf{n}\cdot(\mathbf{D}\cdot\mathbf{n})\right]\mathbf{I}}{\left|\mathbf{P}^{\text{n}}(\mathbf{n})\cdot\mathbf{D}\cdot\mathbf{n}\right|}\cdot\dot{\mathbf{n}}(\mathbf{n})\right)
\label{eq:df_veer}
\end{multline}
\par
The plane-strain DEM data reveals behavior
that is similar to the triaxial simulations that were discussed earlier.
The twirling rate 
$(\partial\widehat{\mathbf{f}^{\text{t}}}
(\mathbf{n})/\partial t)_{\textrm{twirl}}$
for plane-strain biaxial compression is also directed roughly opposite
$\dot{\mathbf{n}}$, but with one difference. Although the twirling
rate and $\dot{\mathbf{n}}(\mathbf{n})$ are roughly opposed in the
biaxial simulations, they are not perfectly counter-aligned: the DEM
data show that the twirling rate consistently includes a small tangential
component that is orthogonal to the $\mathbf{t}(\mathbf{n})$ direction~---
a small component that is aligned with the veering rate given in 
Eq.~(\ref{eq:df_veer}).
With this observation,
we speculate that the original approximation in Eq.~(\ref{eq:ft_twirl})
can be applied to non-triaxial loading by making the following adjustment:
\begin{equation}
\left(\frac{\partial\widehat{\mathbf{f}^{\textrm{t}}}(\mathbf{n})}{\partial t}\right)_{\textrm{twirl}}-\left(\frac{\partial\widehat{\mathbf{f}^{\textrm{t}}}(\mathbf{n})}{\partial t}\right)_{\textrm{veer}}\approx\:-\gamma_{\text{twirl}}\,\widehat{f^{\text{t}}}(\mathbf{n})\,\dot{\mathbf{n}}(\mathbf{n})
\label{eq:twirl_minus_veer}
\end{equation}
Allowing for considerable scatter in the DEM data, this approximation,
with $\gamma_{\text{twirl}}=1.0$, gives a reasonable fit to the DEM
data of both triaxial and plane-strain simulations.
\subsubsection{Rate of tangential force at the critical state}
In a previous section, DEM simulations were used to extract an approximation
of the material rate of normal force during sustained deformation
at the critical state (Section~\ref{sub:Normal_large}). The process
was aided by the use of Eq.~(\ref{eq:rate_fn}) and by the presumption
of a vanishing rate $\left.\partial\widehat{\mathbf{f}^{\text{n}}}/\partial t\right|_{\text{n}}$
during steady, critical state loading. We now apply a similar approach
to tangential force, by using Eq.~(\ref{eq:rate_ft}) to find the
tangential vector rate $(\partial\widehat{\mathbf{f}^{\text{t}}}(\mathbf{n})/\partial t)_{\text{matl}}$.
Because the other terms in Eq.~(\ref{eq:rate_ft}) are aligned
with the migration field $\dot{\mathbf{n}}(\mathbf{n})$, the unit
direction $\mathbf{s}(\mathbf{n})$ of the tangential material rate
must also be aligned with $\dot{\mathbf{n}}(\mathbf{n})$. The DEM
experiments show that at certain orientations, the directions $\mathbf{s}(\mathbf{n})$
and $\dot{\mathbf{n}}(\mathbf{n})$ do indeed coincide; however, at
other orientations, the simulations show that they are collinear but
in \emph{opposite} directions. This paradoxical observation is reconciled
by considering another observation: at those orientations $\mathbf{n}$
where the particles approach each other~--- when the normal rate
$(\partial\widehat{f^{\text{n}}}(\mathbf{n})/\partial t)_{\textrm{matl}}$
is compressive~--- the direction of the tangential rate
coincides with that of $\dot{\mathbf{n}}(\mathbf{n})$. Contrarily,
for orientations $\mathbf{n}$ at which particles tend to withdraw
from each other, the tangential rate and $\dot{\mathbf{n}}(\mathbf{n})$
are counter-aligned. These observations resemble the behavior of
a frictional block system that is loaded with normal and tangential
forces, such as that described by \citet{Bazant:1991a}, \S10.7. The
DEM simulations also show that the magnitude of the tangential material
force rate
correlates with the magnitude of the migration rate, 
$|\dot{\mathbf{n}}(\mathbf{n})|$:
the material rate is larger at orientations where the particles are
migrating more rapidly across each other. When considered together,
these observations suggest the following form for the tangential material
rate at large strains:
\begin{equation}
\left(\frac{\partial\widehat{f^{\text{t}}}(\mathbf{n})}{\partial t}\right)_{\mathrm{matl}}\mathbf{s}(\mathbf{n})\,\approx\,\zeta^{\text{t}}\left(\frac{\partial\widehat{f^{\text{n}}}(\mathbf{n})}{\partial t}\right)_{\mathrm{matl}}\frac{\left|\dot{\mathbf{n}}(\mathbf{n})\right|}{\dot{\epsilon}_{\mathrm{oct}}}\,\mathbf{t}(\mathbf{n})
\label{eq:ft_rate_matl}
\end{equation}
Although Eq.~(\ref{eq:ft_rate_matl}) is consistent with a frictional
system, the factor $\zeta^{\text{t}}$ was measured as only 0.067
in the simulations, a value much smaller than the 0.50 
friction coefficient between particles.
A larger factor might have been realized
if frictional sliding were to occur at all contacts, but at the critical
state, the DEM simulations show that only 18\% of contacts
are sliding (see also \citealt{Thornton:2000a}), as particles tend
to roll rather than slide at their contacts (\citealt{Kuhn:2004k}).
\par
\begin{figure}
\begin{centering}
\includegraphics{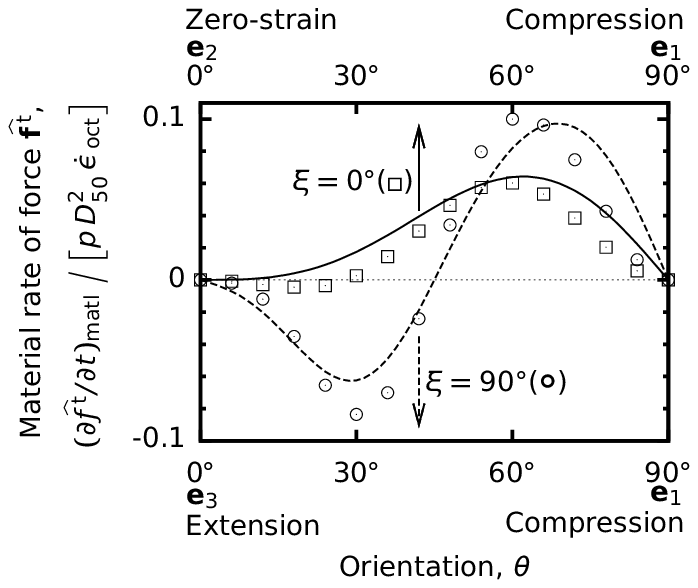}
\par\end{centering}
\caption{Material rates of tangential force
creation during plane-strain biaxial compression at the critical state: data
from DEM simulations (symbols), and rates from Eq.~(\ref{eq:ft_rate_matl})
with $\zeta^{\text{t}}=0.067$ (lines), 
along two $\xi$ meridians (Fig.~\ref{fig:migration}). }
\label{fig:Material_rate_ft_Steady}
\end{figure}

Equation~(\ref{eq:ft_rate_matl}) is compared with data from DEM
simulations in Fig.~\ref{fig:Material_rate_ft_Steady}, which shows
results along two meridians of the unit sphere. Equation~(\ref{eq:ft_rate_matl})
is in general agreement with the experimental data.%
\subsection{\label{section:contact_matl_rate}Rate of contact creation at the
critical state}
The material rate of contact density, ${(\partial\widehat{g}(\mathbf{n})/\partial t)}_{\mathrm{matl}}$,
is the net rate at which contacts are created or extinguished during
deformation, as in Eq.~(\ref{eq:g_rate}). The DEM simulations show
that contacts are predominantly created at orientations $\mathbf{n}$ 
in which the
deformation $\mathbf{D}$ produces compression between particle pairs,
whereas contacts are predominantly broken at orientations of extension. 
These trends
resemble the situation with normal force density, so we begin by assuming
that the net contact creation rate is proportional to the 
rate at which particles approach (or withdraw from) each other~---
the rate $\partial\ell(\mathbf{n})/\partial t$ in Eq.~(\ref{eq:dl_dt}).
In variance with the force rate, however, we now use the deviatoric
rate $\mathbf{D}^{\prime}$ ($=D_{ij}-D_{kk}\delta_{ij}/3$) instead
of the full rate $\mathbf{D}$, nullifying the influence of bulk volume
change on the rate of contact creation. This modification is justified
by two observations. The contact network among particles is most effectively
rearranged by bulk distortion: although pure isotropic compression
will increase the contact forces, it does not greatly alter the number
or orientations of the contacts. 
Furthermore, the dilation that usually accompanies the
shearing of dense granular materials does not appreciably disengage
contacts: the number of contacts will typically remain nearly constant
during post-peak deformation, even as the material is vigorously dilating
\citep{Thornton:2000a}. 
We should expect, however, that the contact
density rate $(\partial\widehat{g}(\mathbf{n})/\partial t){}_{\mathrm{matl}}$
will depend on the average normal force $\overline{f^{\textrm{n}}}(\mathbf{n})$:
larger contact forces imply greater contact indentations and require
a greater motion $d\ell$ to disengage the contacts. In this regard,
we introduce a reference movement for the Hertz contact between an
$m$th pair of elastic spheres (see Eq.~\ref{eq:k_nm}),
\begin{equation}
\frac{f^{\textrm{n},m}}{k^{\textrm{n},m}}\quad\textrm{or}\quad\left[\frac{\left(f^{\textrm{n},m}\right)^{2}(1-\nu)^{2}}{3G^{2}R^{m}}\right]^{1/3}
\label{eq:fn_over_kn}
\end{equation}
This reference movement would cause two spheres, initially pressed
together with force $f^{\textrm{n},m}$, to withdraw and disengage,
if the original stiffness $k^{\textrm{n},m}$ was active throughout
the withdrawal process. The material rate of contact density is approximated
as the particle withdrawal rate $\partial\ell(\mathbf{n})/\partial t$
divided by the average reference movement $\overline{f^{\textrm{n}}}(\mathbf{n})/\overline{k^{\text{n}}}$
and multiplied by the current density $\widehat{g}(\mathbf{n})$.
Applying Eqs.~(\ref{eq:avg_f}), (\ref{eq:dl_dt}), and~(\ref{eq:kn_hat}),%
\begin{equation}
\left(\frac{\partial\widehat{g}(\mathbf{n})}{\partial t}\right)_{\mathrm{matl}}\approx-\beta^{\text{n}}\,\widehat{k^{\text{n}}}(\mathbf{n})\frac{\widehat{g}(\mathbf{n})}{\widehat{f^{\text{n}}}(\mathbf{n})}\,\overline{\ell}\,\left[\mathbf{n}\cdot(\mathbf{D}^{\prime}\cdot\mathbf{n})\right]
\label{eq:g_rate_malt}
\end{equation}
where $\mathbf{D}^{\prime}$ is used in place of $\mathbf{D}$. This
material rate is negative at orientations $\mathbf{n}$ where particles
withdraw from each other~--- when $\mathbf{n}\cdot(\mathbf{D}^{\prime}\cdot\mathbf{n})$
is positive (tensile).
\par
\begin{figure}
\begin{centering}
\includegraphics{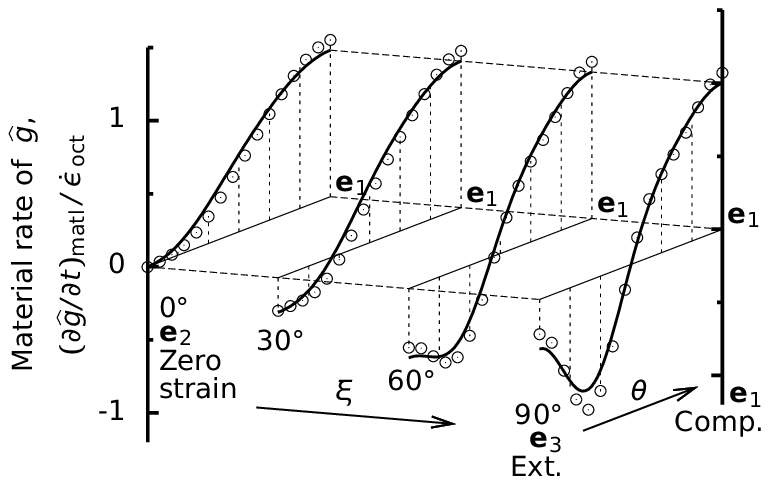}
\par\end{centering}
\caption{Material rates of contact creation during 
plane-strain
biaxial compression at the critical state: data from DEM simulations
(symbols~$\circ$) and rates
from Eqs.~(\ref{eq:g_rate_malt}) and~(\ref{eq:beta_B})
with $B^{\text{n}}=6.3$ (lines~---).}
\label{fig:g_rate}
\end{figure}

Figure~\ref{fig:g_rate} compares Eq.~(\ref{eq:g_rate_malt}) with
DEM data of biaxial plane-strain compression at the critical state, 
showing that the equation closely fits the data. 
\par
If we compare the material rates of contact force creation
and of contact creation (Eqs.~\ref{eq:fn_rate_matl2} and~\ref{eq:g_rate_malt}),
we see that the material rate of force is proportional to $\widehat{g}^{2/3}\widehat{f^{\text{n}}}^{1/3}$;
whereas the material rate of contact creation is proportional to $\widehat{g}^{5/3}\widehat{f^{\text{n}}}^{-2/3}$.
Despite the different scaling of these two rates, the DEM simulations
reveal that the same $\beta^{\text{n}}$ value applies to both rates:
the generation of contacts and the generation of contact force 
apparently share a common origin.
\subsection{\label{section:Diffusion_rates}Diffusion rate at the critical state}
Classical diffusion theory explains the diffusion of a molecular species,
either within itself (self-diffusion) or through other molecular species
(e.g., \citealp{Jeans:1962a}). The process is driven by random fluctuations
among the molecules' velocities, displacing them from their original
positions at time $t=0$. With these random motions, individual displacements
$\delta\mathbf{r}^{m}(t)$ increase with time, such that the collective
mean-square displacement is roughly proportional to time:
\begin{equation}
\left\langle \delta r_{1}^{2}\right\rangle +\left\langle \delta r_{2}^{2}\right\rangle +\ldots+\left\langle \delta r_{d}^{2}\right\rangle =2dDt
\label{eq:DiffusionRate}
\end{equation}
In this form of Einstein-Smoluchowski diffusion, $d$ is the spatial
dimension; the $\delta r_{j}$ are the separate
$j$-components of the randomly advancing displacements; and the diffusion
coefficient $D$ is a measure of the time rate of these growing displacements.
In our application, the displacements $\delta\mathbf{r}^{m}$ are
not of molecules or particles; rather, they are the tangential, angular
displacements of individual contact orientations $\delta\dot{\mathbf{n}}^{m}$
on the unit sphere~--- the fluctuations in Eq.~(\ref{eq:Fluctuations})~---
that occur as particles slide or roll across each other during bulk
deformation.
\par
Equation~(\ref{eq:DiffusionRate}) provides the means for experimentally
measuring a diffusion coefficient~--- in particular, the coefficient
$D_{g}$ of contact self-diffusion in 
Eqs.~(\ref{eq:ContactDiffusion})--(\ref{eq:ft_diffusion}). %
\begin{figure*}
\begin{centering}
\mbox{%
\subfigure[]{\includegraphics{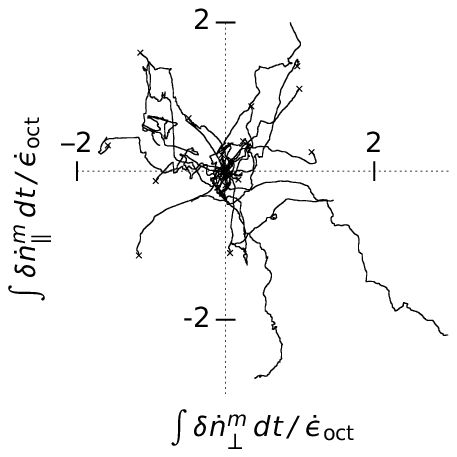}}%
\quad\quad%
\subfigure[]{\includegraphics{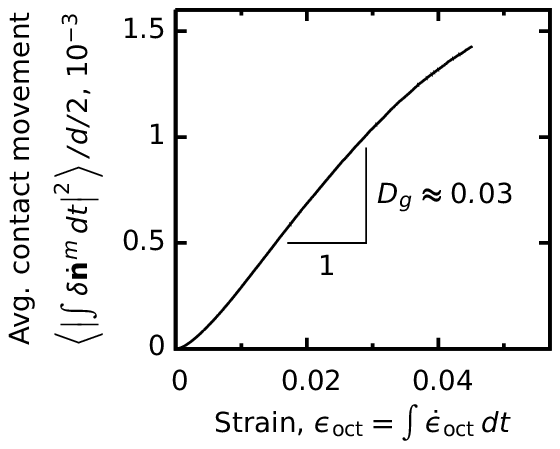}}%
}%
\par\end{centering}
\caption{Fluctuations of contact motions at the critical
state: (a)~movements of twenty-five contacts on the unit sphere and
(b)~estimation of the diffusion coefficient 
$D_{g}$ (\emph{cf.} Eq.~\ref{eq:DiffusionRate}).}
\label{fig:Diffusion}
\end{figure*}
The tangential movements of several contacts are plotted in 
Fig.~\ref{fig:Diffusion}a
over the course of 0.045 strain at the critical state. The figure
illustrates the erratic, zigzag nature of contact movement (i.e.,
a small mean free path). 
To measure the coefficient of contact diffusion $D_{g}$, 
we tracked the long-term movements, 
$\int\delta\dot{\mathbf{n}}^{m}\, dt$,
of over 150,000 contacts as the assembly was being deformed at the
critical state. 
The mean-square cumulative contact displacements are
plotted in Fig.~\ref{fig:Diffusion}b as a function of the advancing
octahedral strain $\epsilon_{\mathrm{oct}}$. The result is a nearly
linear relation, consistent with the conceptual Eq.~(\ref{eq:DiffusionRate}),
with cumulative strain replacing time.
The slope of 0.03 is the contact diffusion coefficient $D_{g}$ that
appears in Eqs.~(\ref{eq:ContactDiffusion})--(\ref{eq:ft_diffusion}),
having units of radians$^{2}$ per unit of strain $\epsilon_{\mathrm{oct}}$.
This value of $D_{g}$ is fairly small when compared with the material
rate of $\widehat{g}$ shown in Fig.~\ref{fig:g_rate}.
\citet{Didwania:2001a} also measured a relatively small rate for a
\emph{kinematic} diffusion derived from the relative translational
velocities between neighboring (but not necessarily contacting)
particle pairs.
\section{\label{sec:pi_plane}Effect of the intermediate principal stress}
The conventional measure of soil strength is the friction angle $\phi$,
based upon the major and minor principal stresses at failure:
$\phi=\sin^{-1}(\sigma_{1}-\sigma_{3})/(\sigma_{1}+\sigma_{3})$.
Tests using advanced true-triaxial and hollow ring torsion apparatus
demonstrate that strength depends on the intermediate principal stress
$\sigma_{2}$, whose relative magnitude is usually represented by
the $b$-value, 
defined as \mbox{$b=(\sigma_{2}-\sigma_{3})/(\sigma_{1}-\sigma_{3})$}.
Certain phenomena, however, can obscure the influence of the intermediate
stress in laboratory tests. Soils exhibit a propensity for inhomogeneous
deformation in the form of shear bands. Shear bands usually appear
near the peak stress, and their emergence can 
alter the subsequent stress-strain behavior
and the measured strength. The emergence
of shear bands can be either suppressed or promoted by the particular
specimen dimensions and boundary conditions, so that the measured
influence of the intermediate principal stress is subject to the vagaries
of the testing equipment (see \citealt{Lade:2006a} for a review).
\par
DEM simulations were conducted with cubical assemblies measuring about
13.4 particle diameters between periodic boundaries 
(Section~\ref{section:DEM}).
Although deformation within an assembly is non-uniform, large-scale
localization, such as shear banding, is unable to develop within such
limited assembly dimensions. The behavior observed in the simulations
can, therefore, be considered close to the underlying material behavior
that would prevail during homogeneous deformation.
\par
\begin{figure*}
\begin{centering}
\mbox{%
\subfigure[]{\includegraphics{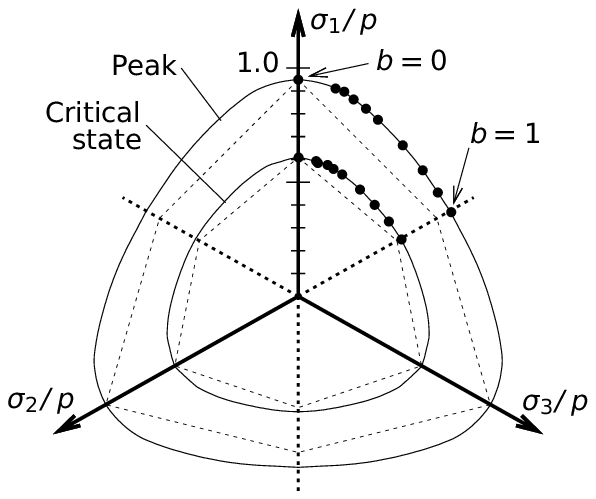}}%
\quad%
\subfigure[]{\includegraphics{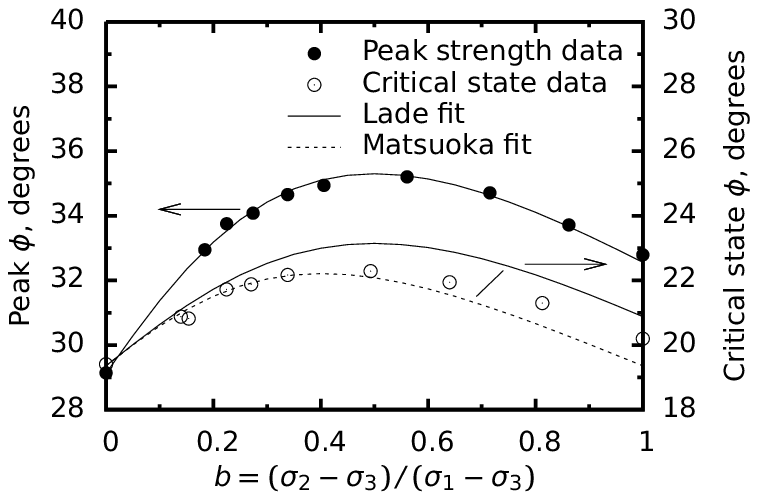}}%
}
\par\end{centering}
\caption{Results of DEM simulations of critical 
state flow with different intermediate principal stresses.}
\label{fig:Results_b_DEM}
\end{figure*}

Figure~\ref{fig:Results_b_DEM} shows results of these DEM simulations,
conducted with different intermediate stresses $\sigma_{2}$.
Ten sets of simulations were conducted: triaxial compression ($b=0$),
triaxial extension ($b=1$), and eight sets with intermediate conditions.
Each set involved loading the same twenty randomly generated assemblies
and averaging the results (see Section~\ref{section:DEM}). 
During loading, mixed boundary conditions were applied in the following manner.
A constant lateral stress $\sigma_{3}$ was maintained while the specimens
were compressed at a constant rate in the $x_{1}$ direction ($\dot{\epsilon}_{11}=\text{constant}$).
Strain control was also maintained in the intermediate direction,
with $\epsilon_{22}$ advanced in fixed proportions of $\epsilon_{11}$
(i.e., engineering strain rates $\dot{\epsilon}_{22}=a\dot{\epsilon}_{11}$,
with the constant $a=-0.5,$ $-$0.375, $-$0.25, $-$0.125, 0, 0.25,
0.5, 0.75, and 1). 
Tests were stopped well after the critical state was
attained and the $b$-value had become stationary, usually at a compressive
engineering strain $-\epsilon_{11}$ of~$0.30$ to~$0.35$.
\par
Figure~\ref{fig:Results_b_DEM}a shows the failure conditions at
the peak and critical states. The principal stresses are plotted on
the deviatoric pi-plane and are normalized by dividing by the current
mean stress $p$. 
Each smooth, curved envelope is a spline fit of ten data points.
Because the initial fabric was isotropic, an entire
failure envelope is symmetrically produced from the ten results that
are shown as dots in the first sextant. 
Flattened Tresca hexagons
are also shown, since these envelopes would apply if strength were
independent of the intermediate principal stress. 
Neither the peak nor
critical state strengths agree with this idealized condition. 
The critical state results, however, form a slightly less rotund envelope, 
lying closer to the Tresca condition.
\par
The results are shown in more detail in Fig.~\ref{fig:Results_b_DEM}b,
which gives the friction angles $\phi$ for the ten $b$-values.
The figure also shows two commonly used methods for fitting a strength
envelope to experimental soil data: 
the Lade and Matsuoka methods \citep{Lade:1975a,Matsuoka:1976a},
which have both been calibrated to the triaxial compression strengths
($b=0$). The Lade method closely fits the DEM strength data at the
peak state, although neither fitting method captures strength at the
critical state.
\par
The micro-mechanical model in 
Sections~\ref{sec:Theory} and~\ref{sec:Quantifying}
can be used for predicting strength at the critical state.
The three Eqs.~(\ref{eq:g_rate}), (\ref{eq:rate_fn}), and~(\ref{eq:rate_ft})
are general rate expressions for the densities 
$\widehat{g}(\mathbf{n})$,
$\widehat{\mathbf{f}^{\text{n}}}(\mathbf{n})$,
and $\widehat{\mathbf{f}^{\text{t}}}(\mathbf{n})$. 
At the critical state, the three densities
are stationary, and the total rates are zero:
\begin{equation}
\left.\frac{\partial\widehat{g}(\mathbf{n})}{\partial t}\right|_{\mathbf{n}}=0\:,\quad\left.\frac{\partial\widehat{\mathbf{f}^{\text{n}}}(\mathbf{n})}{\partial t}\right|_{\mathbf{n}}=0\:,\quad\left.\frac{\partial\widehat{\mathbf{f}^{\text{t}}}(\mathbf{n})}{\partial t}\right|_{\mathbf{n}}=0
\label{eq:Rates_equal_zero}
\end{equation}
Each of these differential equations can be expanded and expressed with the
functional forms that were developed in 
Section~\ref{sec:Quantifying} (e.g. $\dot{\mathbf{n}}(\mathbf{n})$, 
$(\partial\widehat{f^{\text{n}}}(\mathbf{n})/\partial t)_{\text{matl}}$,
etc.). 
In principle, these three equations can be solved for the scalar density
functions $\widehat{g}(\mathbf{n})$, $\widehat{f^{\text{n}}}(\mathbf{n})$,
and $\widehat{f^{\text{t}}}(\mathbf{n})$, which then can be used
in Eqs.~(\ref{eq:fn_plus_ft}) and~(\ref{eq:Stress3}) to find the
stress tensor at the critical state. 
Equations~(\ref{eq:Rates_equal_zero}) are complex: 
they are coupled non-linear partial differential equations on the
surface of the unit sphere.
Looking beyond this difficulty, each of the three equations involve
the deformation rate $\mathbf{D}$, which can be considered an input
parameter. The solution of the three equations and the resulting stress
tensor will, therefore, depend upon the direction of deformation.
\par
Equations~(\ref{eq:Rates_equal_zero}) were solved for several constant-volume
deformation rates $\mathbf{D}$, as would apply at the critical state
($D_{kk}=0$). The coupled non-linear equations were solved using
the method of weighted residuals, by approximating each density function
as a series of even-powered spherical harmonics, for example,
\begin{equation}
\widehat{g}_{\text{A}}(\mathbf{n})=\!\!\!\sum_{\gamma=2,4,6,\ldots}\,\sum_{i_{1},i_{2},\ldots i_{\gamma}}\!\!\!\widehat{G}_{i_{1}i_{2}i_{3}\ldots i_{\gamma}}n_{i_{1}}n_{i_{2}}n_{i_{3}}\ldots n_{i_{\gamma}}
\notag
\end{equation}
where the $\widehat{G}_{\cdots}$ are scalar coefficients, 
and where the direction index $i_{j}\in\{1,2,3\}$. 
Approximations for $\widehat{g}(\mathbf{n})$,
$\widehat{f^{\text{n}}}(\mathbf{n})$, 
and $\widehat{f^{\text{t}}}(\mathbf{n})$
were substituted into the appropriate forms of all terms
on the right sides of expressions~(\ref{eq:g_rate}),
(\ref{eq:rate_fn}), and~(\ref{eq:rate_ft}), and the harmonic coefficients
were sought to minimize these expressions over the unit sphere:
\begin{equation}
\text{min. }\!\!\!
\int_{\Omega}
\left[\left(\frac{\partial\widehat{g}_{\text{A}}}{\partial t}\right)^{2}
\!\!+\left(\frac{\partial\widehat{f^{\text{n}}}_{\text{A}}}{\partial t}\right)^{2}
\!\!+\left(\frac{\partial\widehat{f^{\text{t}}}_{\text{A}}}{\partial t}\right)^{2}\right]d\Omega
\notag
\end{equation}
thus approximating a simultaneous solution of all three 
Eqs.~(\ref{eq:Rates_equal_zero}).
\par
\begin{figure}
\centering
\includegraphics{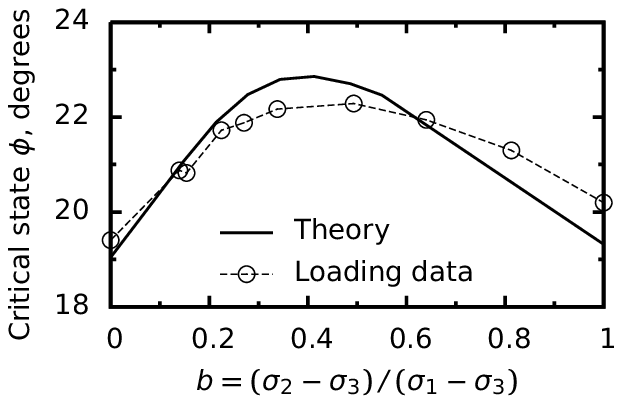}
\caption{Effect of the intermediate principal stress
on strength: comparison of micro-mechanical theory with DEM simulations.}
\label{fig:Phi_vs_b_model}
\end{figure}

The solution results are shown in Fig.~\ref{fig:Phi_vs_b_model}.
The differential equations were quantified with the particle properties
$E$ and $\nu$ and with the micro-mechanical transport properties
that were measured at the critical state: $\alpha$, $B^{\text{n}}$,
$\beta_{2}^{\text{n}}$, $\gamma_{\text{twirl}}$, $\zeta^{\text{t}}$,
and~$D_{g}$. 
These values had been extracted from the single set of DEM simulations
of plane-strain biaxial loading ($a=0$), and we simply used the same values
in solving the differential equations 
for other deformation directions~$\mathbf{D}$.
A common mean stress and average coordination number were imposed
as auxiliary constraints ($p=490$~kPa, $2M/N=4.0$). 
The solutions of Eqs.~(\ref{eq:Rates_equal_zero}) imply stress tensors that
would, in theory, produce critical state flow under various deformation
directions $\mathbf{D}$. Figure~\ref{fig:Phi_vs_b_model} shows
that these solutions compare reasonably well with data from the ten
sets of true-triaxial simulation experiments.%
\section{Conclusion}
The Paper presents a model for the evolution of fabric and stress
in granular materials. A distinction is made between evolution effects
produced by the interactions of particles~--- termed {}``material''
effects~--- and effects that are due to an \emph{en masse} shifting
of contacts and forces from one orientation to another. The latter
include convection, diffusion, and twirling effects. The DEM simulations
show that the material rate has a consistent hardening influence during
loading: by itself, the material rate increases the deviatoric stress
and induces fabric anisotropy. The other rates have a consistent softening
influence, reducing the anisotropies of contact force and contact
orientation. The model is most useful in assessing failure at large
strains, when the two competing changes are of roughly equal magnitude
and both must be tracked during the loading process. At the critical
state, the effects are in balance, producing stationary stress and
fabric. With this observation, the Paper develops one application
of the model: a prediction of the effect of the intermediate principal
stress on strength at the critical state. 
\par
Although the model's predictions compare favorably with DEM simulations
of small assemblies of 4096 particles, such results may seem questionable
when applied to the scale of problems that are encountered in industrial
and geotechnical situations. In these large-scale problems, deformation
and failure are not homogeneous, but are instead concentrated within
shear bands and other localization zones. The Paper's model is based
upon an averaging of trends among the many contacts within a small
representative volume, and this averaged behavior can be thought
to apply to continuum points or to small regions within a shear band,
rather than encompassing the entire band thickness. By pursuing a
point-based continuum view, the model might be used to develop advanced,
comprehensive continuum models that incorporate a length scale and
can predict the emergence and evolution of a shear band. Fulfilling
this promise requires work beyond that of the Paper. The effect of
the intermediate principal stress, suitably predicted by the model, 
is but one phenomenon that is known
to influence the possibility and nature of shear bands. 
At least three other phenomena are also known to be relevant, 
and these phenomena
are listed as possible future applications of the model, which might
culminate in a more comprehensive continuum description.
\begin{enumerate}
\item%
As shearing progresses within a granular material, shear bands begin
to form before the peak stress is attained, but these early bands
are usually transitory and will form, grow, and then disappear.
Once the peak stress is reached and the soil begins to soften, the
numerous bands coalesce into a few persistent shear bands~--- perhaps
a single shear band~--- and further deformation becomes concentrated
and captive within these persistent features. The theoretical framework
in the Paper could be used to model the onset and evolution of softening.
The divergence, convection, and diffusion processes reduce the pronounced
anisotropies of fabric and force that are attained at the peak state,
which might account, at least partially, for the strength softening.
\item%
Incremental strains within a shear band are often not aligned with
the stress increments: granular materials exhibit a non-coaxiality
of the principal strain and stress increments. 
Such non-coaxiality
can favor an abrupt change in the direction of deformation, as at
the start of shear banding. 
The non-coaxiality of stress and strain
increments is likely due to induced fabric and force anisotropies,
which favor stress increments in certain directions, regardless
of the direction of deformation. 
The model is naturally suited to account for such effects of
fabric anisotropy.
\item%
Shear bands have a characteristic thickness that is related to particle
size. A comprehensive and verifiable explanation of this characteristic
remains an open problem in granular mechanics. Several continuum theories
have been proposed as a rationale for band thickness, but to the author's
knowledge, only one such explanation has been confirmed: a dependence
of the stress increment upon the spatial gradients of strain \citep{Kuhn:2002a}.
The model in the Paper can, perhaps, be extended to understand and
explain the effect of strain gradients, since such gradients will
alter the contact migration pattern of Eqs.~(\ref{eq:n_align}) 
and~(\ref{eq:ndot_alpha}),
changing the evolution of fabric and stress. 
\end{enumerate}
An application of the model to these three phenomena will likely require
an understanding of density rates during unloading, a matter not addressed
in the Paper. Although extending the model to include unloading rates
presents a further challenge, once achieved, the model might also
provide a micro-mechanical alternative to the notion of a yield surface.
\bibliographystyle{MofM}
%
%

%
%
\end{document}